%% file: corlep.tex
\documentclass[12pt]{article}
\usepackage{a4wide}
\usepackage{epsfig}
\usepackage{bbm}

\newcommand{\oone}{\hbox{$1\kern-2.5pt\hbox{\rm l}$}}
\newcommand{\ssigma}{\hbox{$\kern2.5pt\vrule height4.5pt\kern-2.5pt\sigma$}}
\newcommand{\complex}{\hbox{\kern2.5pt\vrule height8.5pt\kern-2.5pt C}}

\newcommand{\GeV}{{\rm\,GeV}}
\newcommand{\pfrac}[2]{\left(\frac{#1}{#2}\right)}

\newcommand{\Li}{{\rm Li}}
\newcommand{\imag}{\mathop{\rm Im}\nolimits}
\newcommand{\real}{\mathop{\rm Re}\nolimits}
\newcommand{\eps}{\varepsilon}
\newcommand{\sxi}{{\textstyle\sqrt\xi}}
\newcommand{\Tr}{\mathop{\rm Tr}\nolimits}
\newcommand{\slp}{p\kern-5pt/}
\newcommand{\sls}{s\kern-5pt/}
\newcommand{\nl}{\nonumber\\&&\strut}

\newcommand{\tintsm}{\left((2-\xi)\left(\hat I^{L_0}_{t-}(0)
  -I^{L_0}_{t+}(0)\right)-4v\ell_0^-\right)}
\newcommand{\tintsp}{\left((2-\xi)\left(\hat I^{L_0}_{t-}(0)
  +I^{L_0}_{t+}(0)\right)-4v(\ell_0^-+\ell_0^+)\right)}
\newcommand{\tintpm}{\left(I^{L_+}_{t-}(0)-I^{L_+}_{t+}(0)\right)}
\newcommand{\tintpp}{\left(I^{L_+}_{t-}(0)+I^{L_+}_{t+}(0)\right)}

\begin{document}

\thispagestyle{empty} 
\begin{flushright}
MITP/16-057\\
\end{flushright}
\vspace{0.5cm}

\begin{center}
  {\Large\bf Analytical $O(\alpha_s)$ corrections to the beam frame\\[.3cm]
  double-spin density matrix elements of $e^+e^-\to t\bar t$}\\[1.3cm]
{\large L.~Kaldam\"ae$^1$, S.~Groote$^1$ and J.G.~K\"orner$^2$}\\[1cm]
$^1$ Loodus- ja Tehnoloogiateaduskond, F\"u\"usika Instituut,\\[.2cm]
  Tartu \"Ulikool, W.~Ostwaldi 1, 50411 Tartu, Estonia\\[7pt]
$^2$ PRISMA Cluster of Excellence, Institut f\"ur Physik,
  Johannes-Gutenberg-Universit\"at,\\[.2cm]
  Staudinger Weg 7, 55099 Mainz, Germany
\end{center}

\vspace{1cm}
\begin{abstract}\noindent
We provide analytical results for the $O(\alpha_s)$ corrections to the
double-spin density matrix elements in the reaction $e^+e^-\to t\bar t$. These
concern the elements $ll$, $lt$, $ln$, $tt$, $tn$, and $nn$ of the double-spin
density matrix elements where $l,t,n$ stand for longitudinal, transverse and
normal orientations with respect to the beam frame spanned by the electron and
the top quark momentum.
\end{abstract}

\newpage

\section{Introduction}
The measurement of polarisation effects in top and antitop quark events
produced at $e^+e^-$ colliders is very interesting in that one can test the
details of the Standard Model production and decay mechanisms of the produced
top--antitop quark pairs and their decays. In addition such measurements can
be used to constrain deviations from Standard Model
couplings~\cite{Peskin:1992,Ladinsky:1992vv,Barklow:1994,Schmidt:1995mr}.
While single-spin polarisation effects are not observed for hadronically
produced top quark pairs as at the LHC due to parity
conservation~\cite{Korner:2003zq}, single-spin polarisation effects are
present in $e^+e^-$ colliders due to the existence of parity-violating
components in the production mechanism~\cite{Mahlon:1995zn}. Parity
conservation in the production mechanism for the top and antitop pair,
however, allows for spin--spin correlation effects between top and antitop
quark spins as at the LHC. The analysis of spin--spin correlations of top
quark pairs has become a very popular subject in the last few years (see e.g.\
Ref.~\cite{ATLAS:2012ao}).

The top quark retains its polarisation at birth when it decays because its
lifetime is so short that it decays before hadronisation can wash out its
polarisation. Therefore, polarisation effects at the envisaged $e^+e^-$
colliders ILC and CLIC should help to find new physics. Of interest is also
the role of quark mass effects in the production of quarks and gluons in
$e^+e^-$ annihilations. Analytical results for the $O(\alpha_s)$ radiative
corrections to longitudinal single-spin polarisation including quark mass
effects can be found in Refs.~\cite{Korner:1993dy,Tung:1994fe,Tung:1996dq,%
Groote:1995yc,Groote:1996nc}, and corresponding results for the transverse and
normal polarisation components can be found in Ref.~\cite{Groote:1995ky}. In
Refs.~\cite{Tung:1997ur,Groote:1997su,Groote:2009zk} analytical results for
the $O(\alpha_s)$ radiative corrections to longitudinal spin--spin
correlations between massive quark pairs can be found.

The aim of this paper is twofold. On the one hand we provide an independent
check of the numerical next-to-leading-order (NLO) results presented in
Refs.~\cite{Brandenburg:1998xw,Brandenburg:1999ss}. On the other hand we
provide analytical results for the $O(\alpha_s)$ radiative corrections to the
transverse and transverse normal spin--spin correlation asymmetry (called
transverse and normal spin--spin asymmetry for short) and its polar angle
dependence for massive quark pairs produced in $e^+e^-$ annihilations. 

To define and measure single-spin and spin--spin correlation observables
requires the definition of one or two coordinate systems in the top quark and
antitop quark rest systems. Several coordinate bases have been proposed in
Ref~\cite{Parke:1996pr}, namely the helicity basis, beamline basis and
off-diagonal basis. As advised in Refs.~\cite{Brandenburg:1998xw,%
Brandenburg:1999ss}, in this paper we use a common basis for both top and
antitop quark observables. As the $z$ direction lies along the direction of
the top quark, for the top quark this basis can be referred to as the helicity
basis. However, two linearly independent directions are necessary to build a
frame. The beam frame used here is spanned by the momenta of the electron and
quark. Leading-order results for spin--spin correlations in this frame have
been given in Ref.~\cite{Parke:1996pr} while numerical results for the
$O(\alpha_s)$ radiative corrections can be found in
Refs.~\cite{Brandenburg:1998xw,Brandenburg:1999ss}. However, it is not really
necessary to choose a common reference frame for the top and antitop rest
frames. For example, in Refs.~\cite{Groote:1997su,Groote:2009zk,Kodaira:1998gt}
the respective helicity systems for the top and antitop were chosen which are
not necessarily anticollinear at NLO. Still, the differences are marginal
because the mean deviation from the anticollinearity due to $O(\alpha_s)$
corrections is small~\cite{Groote:1998xc}.

In performing these calculations in the beam frame spanned by the momenta of
the electron and the quark, it turned out that the integrals necessary for
the phase space integration are quite similar to those used by us in former
calculations~\cite{Korner:1993dy,Tung:1994fe,Tung:1996dq,Groote:1995yc,%
Groote:1996nc,Groote:1995ky,Tung:1997ur,Groote:1997su,Groote:2009zk} (see also
Ref.~\cite{Tung:2004ge} for the mathematical background). Calculations
performed in the event frame spanned by the top, antitop and gluon including
also elliptic integrals will be found in a separate
publication~\cite{Kaldamae:2016}.

In the course of this paper we also explain how to measure the spin--spin
correlation (see e.g.\ Refs.~\cite{Grzadkowski:2000nx,Khiem:2015ofa}). As in
the corresponding hadronic case~\cite{Bernreuther:2004jv,Bernreuther:2001rq,%
Bernreuther:2010ny,Bernreuther:2015yna} we investigate spin--spin correlation
effects through
\begin{enumerate}
\item
the double angle distribution\footnote{For hadronically produced top--antitop
quark pairs the single angle coefficients $B_1$ and $B_2$ vanish at LO due to
parity conservation of the strong interactions.}
\begin{equation}
  \frac1\sigma\frac{d\sigma}{d\,\cos\theta_1d\,\cos\theta_2}
  =\frac14\left(1+B_1\cos\theta_1-B_2\cos\theta_2
  -C\cos\theta_1\cos\theta_2\right),
\end{equation}
where $\theta_1$ and $\theta_2$ are the angles between a fixed direction given
by the basis used and the direction of flight of the charged lepton in the
rest frames of the top and antitop quark, respectively, and
\item
the opening angle distributions
\begin{equation}
  \frac1\sigma\frac{d\sigma}{d\,\phi}=\frac14\left(1-D\cos\phi\right),
\end{equation}
where $\phi$ is the angle between the directions of flight of the charged
leptons in the rest frames of the top and antitop quark, respectively.
\end{enumerate}
In Refs.~\cite{Bernreuther:2004jv,Bernreuther:2001rq,Bernreuther:2010ny,%
Bernreuther:2015yna} Bernreuther and coworkers dealt with the subject of
top--antitop spin--spin correlations at hadron colliders. They introduced
a variety of spin--spin observables and gave detailed prescriptions how to
measure these observables.

The paper is organised as follows. In Sec.~2 we deal with the double-spin
density matrix, we specify quantisation axes and introduce observables. In
addition, we explain how these observables are related to angular dependences
measured in subsequent particle cascades. In Sec.~3 we introduce our analytical
$O(\alpha_s)$ results which are found in Appendices~A and~B. The dependence of
the observables on the center-of-mass energy, the polar angle and the initial
beam polarisation is discussed in Sec.~4. Sec.~5 contains our conclusions.

\section{The double--spin density matrix}
Polarisation observables are best described in terms of the spin density
matrix. The two spin states of a spin-$1/2$ fermion are denoted by
$|\lambda=\pm 1/2\rangle$. The two spin states are eigenstates of the spin
operator $J_z$ in a given frame. The coefficients of the normalised spin
density matrix $\hat\rho$ are given by superpositions of the elements
$|\lambda\rangle\langle\lambda'|$. In a moving frame one has the completeness
relation
\begin{equation}
\slp+m_f=\sum_{\lambda=\pm 1/2}|\lambda\rangle\langle\lambda|.
\end{equation}
One can represent the two spin states as components of a two-dimensional Pauli
spinor which implies that one can parametrise the (normalised) spin density
matrix in terms of Pauli matrices according to\footnote{In order to
distinguish between the Pauli matrix and cross section, we use the symbol
$\ssigma$ for the former.}
\begin{equation}
\hat\rho=(\hat\rho_{\lambda,\lambda'})=\frac12(\oone+P^i\ssigma_i),
\end{equation}
where $\vec P=(P^i)$ is the three-dimensional polarisation vector. The
expectation value of an arbitrary operator ${\cal O}$ is obtained by
calculating the trace, $\langle{\cal O}\rangle=\Tr(\hat\rho{\cal O})$.

Next we introduce the double--spin density matrix which is needed for the
discussion of spin--spin correlation effects. Given a quantisation axis for
the observation of single-spin polarisation and spin--spin correlation effects
in $e^+e^-$ collisions, the $4\times4$ un-normalised double density matrix
$\hat\rho$ is parametrised by expanding the density matrix along outer
products of the standard set of $2\times2$ matrices. One has
\begin{equation}
\label{expansion}
\hat\rho=(\hat\rho_{\lambda^{\phantom{}}_1\lambda^{\phantom{}}_2,
  \lambda'_1\lambda'_2})=
\frac14\left(\rho\,\oone\otimes\oone
  +\rho^{e_1^i}\ssigma_i\otimes\oone
  +\rho^{\bar e_2^j}\oone\otimes\ssigma_j
  +\rho^{e_1^i\bar e_2^j}\ssigma_i\otimes\ssigma_j\right)
\end{equation}
where the outer product symbol $\otimes$ denotes the tensor product between
the spin states of the top and antitop quarks according to $(A\otimes B
)_{\lambda^{\phantom{}}_1\lambda^{\phantom{}}_2,\lambda'_1\lambda'_2}
=A_{\lambda^{\phantom{}}_1\lambda'_1}B_{\lambda^{\phantom{}}_2\lambda'_2}$.
The labels $\lambda^{\phantom{}}_1$ ($\lambda'_1$) and $\lambda^{\phantom{}}_2$
($\lambda'_2$) denote the two spin states of the top and antitop quark,
respectively.

\subsection{Quantisation axes}
Given a quantisation axis for each spin degree of freedom represented by the
two orientation vectors $\vec e_1$ and $\vec{\bar e}_2$, the coefficient
functions $\rho$, $\rho^{e_1^i}$, $\rho^{\bar e_2^j}$ and
$\rho^{e_1^i\bar e_2^j}$ denoting the rate function, the un-normalised
single-spin polarisation components of the top and antitop quarks, and the
double-spin correlation component, respectively, can be projected from
Eq.~(\ref{expansion}) by tracing the appropriate products of $\hat\rho$ with
$\oone\otimes\oone$, $e_1^i\ssigma_i\otimes\oone$,
$\bar e_2^j\oone\otimes\ssigma_j$ or
$e_1^i\bar e_2^j\ssigma_i\otimes\ssigma_j$. One obtains
\begin{eqnarray}\label{uparrow1}
\rho&=&\rho(\uparrow,\uparrow)+\rho(\uparrow,\downarrow)
  +\rho(\downarrow,\uparrow)+\rho(\downarrow,\downarrow),\nonumber\\
\rho^{e_1}&=&\rho(\uparrow,\uparrow)+\rho(\uparrow,\downarrow)
  -\rho(\downarrow,\uparrow)-\rho(\downarrow,\downarrow),\nonumber\\
\rho^{\bar e_2}&=&\rho(\uparrow,\uparrow)-\rho(\uparrow,\downarrow)
  +\rho(\downarrow,\uparrow)-\rho(\downarrow,\downarrow),\nonumber\\
\rho^{e_1\bar e_2}&=&\rho(\uparrow,\uparrow)-\rho(\uparrow,\downarrow)
  -\rho(\downarrow,\uparrow)+\rho(\downarrow,\downarrow).
\end{eqnarray}
where $\uparrow$ and $\downarrow$ represent the two orientations with respect
to the quantisation axes. For the production process $e^+e^-\to t\bar t(G)$,
the contributions on the right-hand side of Eq.~(\ref{uparrow1}) are given by
\begin{equation}\label{rho}
\rho(\vec s_1,\vec s_2)=\Tr\left(\hat\rho\,\frac12(\oone+\vec s_1\cdot\ssigma)
  \otimes\frac12(\oone+\vec s_2\cdot\ssigma)\right)=\frac{e^4N_c}{q^4}
  \sum_{i,j=1}^4g_{ij}L^i_{\mu\nu}H^{j\,\mu\nu}=|T_{fi}|^2
\end{equation}
defining the spin dependence of the squared matrix element $|T_{fi}|^2$ of
the production process, where $q$ is the momentum carried by the intermediate
boson ($\gamma$ or $Z$). The multicomponent overall factor $e^4N_cg_{ij}/q^4$
incorporates the boson propagator effect and the multicomponent electroweak
coupling factors $g_{ij}$ which includes the dimensionless part of the
interactions of the fermions with the bosons (cf.\ Ref.~\cite{Groote:1996nc}).
The various components of the lepton tensor $L^i_{\mu\nu}$ and hadron tensor
$H^i_{\mu\nu}$ are decomposed according to
\begin{eqnarray}
L^1=\frac12(L^{VV}+L^{AA})&&L^2=\frac12(L^{VV}-L^{AA})\nonumber\\
L^3=\frac i2(L^{VA}-L^{AV})&&L^4=\frac12(L^{VA}+L^{AV})
\end{eqnarray}
(and accordingly for the $H^i$) where $V$ and $A$ denote the vector and
axial-vector contributions. The double density matrix can now be expanded into
the tensor product of two bases. In Ref.~\cite{Brandenburg:1998xw} the
directions of the top quark ($\hat k$), the electron momentum ($\hat p$) and a
normalised vector ($\hat n$) perpendicular to these two has been used for both
bases. These two bases need not be the same. For convenience of the
phenomenological calculation of the double density matrix via
$\rho(\vec s_1,\vec s_2)$, we choose as bases the bases
$(\vec s_i^{\,T},\vec s_i^{\,N},\vec s_i^{\,L})$ of the two spins $s_i$
($i=1,2$) boosted to the laboratory frame. The respective rest frame spin
vectors are given by
\begin{eqnarray}
\vec s_1^{\,T}=(1,0,0),&&
\vec s_1^{\,N}=(0,1,0),\qquad
\vec s_1^{\,L}=(0,0,1),\nonumber\\[7pt]
\vec s_2^{\,T}=(\cos\theta_{12},0,-\sin\theta_{12}),&&
\vec s_2^{\,N}=(0,1,0),\qquad
\vec s_2^{\,L}=(\sin\theta_{12},0,\cos\theta_{12}),
\end{eqnarray}
where $\theta_{12}$ is the polar angle between the momenta of the top and
antitop quarks. At the Born term level one has $\theta_{12}=\pi$. One obtains
the double expansion
\begin{eqnarray}\label{rhoexp}
\hat\rho&=&\frac14\Big(\rho(\oone\otimes\oone)+\left(\rho^Ts_1^{iT}
  +\rho^Ns_1^{iN}+\rho^Ls_1^{iL}\right)(\ssigma_i\otimes\oone)
  \strut\nonumber\\&&\strut
  +\left(\rho^{\bar T}s_2^{jT}+\rho^{\bar N}s_2^{jN}+\rho^{\bar L}s_2^{jL}
  \right)(\oone\otimes\ssigma_j)\strut\nonumber\\&&\strut
  +\big(\rho^{T\bar T}s_1^{iT}s_2^{jT}+\rho^{T\bar N}s_1^{iT}s_2^{jN}
  +\rho^{T\bar L}s_1^{iT}s_2^{jL}\strut\nonumber\\&&\strut
  +\rho^{N\bar T}s_1^{iN}s_2^{jT}+\rho^{N\bar N}s_1^{iN}s_2^{jN}
  +\rho^{N\bar L}s_1^{iN}s_2^{jL}\strut\nonumber\\&&\strut
  +\rho^{L\bar T}s_1^{iL}s_2^{jT}+\rho^{L\bar N}s_1^{iL}s_2^{jN}
  +\rho^{L\bar L}s_1^{iL}s_2^{jL}\big)(\ssigma_i\otimes\ssigma_j)\Big),\qquad
\end{eqnarray}
The relation between the un-normalised double--spin density matrix and the
differential cross section is given by
\begin{equation}
d\sigma(\vec s_1,\vec s_2)=\frac1{2q^2}\rho(\vec s_1,\vec s_2)dPS.
\end{equation}
For a two-particle final state as in $e^+e^-\to t\bar t$ the Lorentz-invariant
phase space $dPS$ is given by
\begin{equation}
dPS_2=\frac{v}{16\pi}d\cos\theta,
\end{equation}
where $v=\sqrt{1-\xi}$, $\xi=4m^2/q^2$, and $\theta$ is the polar angle
between the momenta of the electron and the top quark. For the three-particle
final state $t\bar tG$ one has two additional integration parameters
$y:=1-2(p_1q)/q^2$ and $z:=1-2(p_2q)/q^2$ which are related to the energies
$E_1=p_1q/\sqrt{q^2}$ and $E_2=p_2q/\sqrt{q^2}$ of the top quark and the
antitop quark in the laboratory frame. In terms of the parameters $y$ and $z$
the three-particle phase space reads
\begin{equation}
dPS_3=\frac{q^2}{32(2\pi)^4}dy\,dz\,d\cos\theta\,d\chi,
\end{equation}
where $\chi$ is the azimuthal angle between the beam plane spanned by the
electron and the top quark, and the event plane spanned by the top quark and
the antitop quark (or gluon).

\subsection{Observables}
We loosely refer to the single-spin polarisation vectors and the spin--spin 
correlation tensors as observables even if the observability of these objects
needs to be specified by e.g.\ their subsequent decay distributions. We shall
return to this point in the next subsection.

Following Refs.~\cite{Brandenburg:1998xw,Brandenburg:1999ss} we define the
observables
\begin{equation}\label{defobs}
O^{e_1e_2}=\frac{d\sigma^{e_1e_2}}{d\sigma},\qquad
O^{e_1}=\frac{d\sigma^{e_1}}{d\sigma},
\end{equation}
where $\vec e_1$ and $\vec e_2$ now are elements of the {\em same\/} frame,
for which, as in Refs.~\cite{Brandenburg:1998xw,Brandenburg:1999ss}, we choose
the top quark rest frame. Differing from Refs.~\cite{Brandenburg:1998xw,%
Brandenburg:1999ss} we use an orthonormal basis in the top quark rest frame.
As in Ref.~\cite{Groote:2010zf} the three orthonormal basis vectors are
defined by
\begin{equation}
\hat t=\frac{(\vec p_{e^-}\times \vec p_t)\times\vec p_t}{|(\vec p_{e^-}
  \times\vec p_t)\times\vec p_t|},\qquad
\hat n=\frac{\vec p_{e^-}\times\vec p_t}{|\vec p_{e^-}\times\vec p_t|},\qquad
\hat l=\frac{\vec p_t}{|\vec p_t|},
\end{equation}
The three orthonormal basis vectors $(\hat t,\hat n,\hat l)$ define our
right-handed orthonormal frame. In order to check on the results of
Refs.~\cite{Brandenburg:1998xw,Brandenburg:1999ss} we have also worked in the
nonorthogonal frame $(\hat k,\hat p,\hat n)$ employed in
Ref.~\cite{Brandenburg:1998xw} where $\hat k=\hat l$ and
$\hat p=\vec p_{e^-}/|\vec p_{e^-}|$. Our unit vector $\hat t$ can be seen to
be a linear superposition of the unit vectors $\hat k$ and $\hat p$ given by
\begin{equation}\label{bfuvector}
\hat t=\frac{\hat k\cos\theta-\hat p}{\sin\theta}.
\end{equation}

Returning to Eq.~(\ref{defobs}), the numerators in Eq.~(\ref{defobs}) are
calculated according to
\begin{eqnarray}\label{rhoee}
d\sigma^{e_1e_2}&=&\frac1{2q^2}\Tr\left(\hat\rho\,e_1^i\frac12\ssigma_i
  \otimes e_2^j\frac12\ssigma_j\right)dPS\ =\ \frac1{2q^2}\rho^{e_1e_2}dPS,\\
d\sigma^{e_1}&=&\frac1{2q^2}\Tr\left(\hat\rho\,e_1^i\frac12\ssigma_i
  \otimes\oone\right)dPS\ =\ \frac1{2q^2}\rho^{e_1}dPS.\label{rhoe}
\end{eqnarray}
The unpolarised rate in the denominators of Eq.~(\ref{defobs}) provides for
the appropriate normalisation and can be calculated according to
\begin{equation}
d\sigma=\frac1{2q^2}\Tr\left(\hat\rho\,\oone\otimes\oone\right)dPS
  \ =\ \frac1{2q^2}\rho\,dPS.
\end{equation}
The phase space element $dPS$ is rather symbolic and stands for a generic
phase space element that remains after single or multiple phase space
integrations.
Summing up events in terms of polarisation degrees according to the three
quantisation axes along $\hat t$, $\hat n$ and $\hat l$, one obtains
contributions to a correlation matrix. This, however, is still not the
quantity observed in the experiment. The polarisation unveils itself rather
by angular distributions of the subsequent decays of the top and antitop
quarks.

\subsection{Polarisation analysis via subsequent cascade decays}
In this subsection we discuss two measurements that allow one to analyse two
particular linear combinations of the spin--spin coefficient functions. These
measurements employ the inclusive semileptonic decays
$t(\uparrow)/\bar t(\uparrow)\to \ell^+/\ell^- +X$ derived from the
dominant decays $t \to b \ell^+ \nu_\ell$ and
$\bar t \to \bar b \ell^- \bar \nu_\ell$ and require the
reconstruction of the momentum directions of the charged leptons in the
respective top/antitop rest frames.

There are two ways to analyse the polarisation of the top quark. The first is
to treat the decay $t\to b+W^+(\to\ell^++\nu)$ as a cascade decay
process~\cite{Fischer:1998gsa,Fischer:2001gp,Czarnecki:2010gb}. The second way
is to analyse the polarised top decay directly in the top quark rest 
frame~\cite{Groote:2006kq}. We shall use the second possibility as has also
been done in Ref.~\cite{Bernreuther:2004jv}. In our theoretical analysis we
work in the narrow-width approximation for the top and antitop quarks which is
well justified since the top quark width is much smaller than its mass. In
order to describe the spin dependence of the cascade decays
$e^+e^-\to t(\to b+\ell^++\nu_\ell)+\bar t(\to\bar b+\ell^-+\bar\nu_\ell)$ we
employ the density matrix formalism of Ref.~\cite{Bernreuther:2004jv}.
The whole cascade process is written in product form in terms of the
production density matrix and the two decay density matrices. The
absolute square of the matrix element for the cascade process including the
spin-density matrices $\hat\rho(t)$ and $\hat\rho(\bar t)$ for the decay of
the top and antitop quark, respectively, is proportional to the trace
\begin{equation}\label{tr3rho}
\Tr\left(\hat\rho(t\bar t)\left(\hat\rho(t)\otimes\hat\rho(\bar t)\right)
  \right)=\hat\rho_{\lambda'_1\lambda^{\phantom\prime}_1}(t)
  \hat\rho_{\lambda^{\phantom\prime}_1\lambda^{\phantom\prime}_2,
  \lambda'_1\lambda'_2}(t\bar t)
  \hat\rho_{\lambda'_2\lambda^{\phantom\prime}_2}(\bar t),
\end{equation}
where $\hat\rho(t\bar t)$ is the double density matrix calculated in this
paper.

The spin-density matrices for the two decaying top and antitop quarks are
given by
\begin{equation}
\hat\rho(t)=\frac{\rho(t)}2\left(\oone+\alpha(t)\hat q_1\cdot\ssigma\right),
  \qquad
\hat\rho(\bar t)=\frac{\rho(\bar t)}2\left(\oone-\alpha(\bar t)\hat q_2\cdot
  \ssigma\right),
\end{equation}
where $\rho(t)$ and $\rho(\bar t)$ are the partial widths of the corresponding
decay channel for the decay of the polarised top and antitop quark,
respectively, and $\alpha(t)$ and $\alpha(\bar t)$ are the analysing powers of
the corresponding decay for the top and antitop spin, respectively, along the
flight directions of the analysing decay particles given by their
three-momentum vector directions $\hat q_1$ and $\hat q_2$ in the rest frames
of the top and antitop quark, respectively. If the decay channels are charge
conjugate to each other as in our case for the top and antitop quarks, the
partial widths and the analysing powers are equal, $\rho(\bar t)=\rho(t)$ and
$\alpha(\bar t)=\alpha(t)$.

The analysing power of the final state lepton has been analysed in
Refs.~\cite{Czarnecki:1990pe,Czarnecki:1993gt,Czarnecki:1994pu} including
radiative corrections. The Born term analysing powers of the three final-state
particles in the decay $\ell^+$, $b$ and $\nu_\ell$ have been summarised in
Table~3 in Ref.~\cite{Korner:1998nc}. The LO spin analysing power in this
decay was found to be $100\%$ if one uses the momentum of the charged lepton
as the analyser. Because the decay $t(\uparrow)\to b\ell^+\nu_\ell$ has the
same flavour structure, the result of Ref.~\cite{Korner:1998nc} can be carried
over to the present case. Let us define unit vectors $\hat q_i$ in the
direction of the charged leptons ($i=1,2$ stand for $\ell^+,\ell^-$). These
unit vectors can be expanded in the frame $(\hat t,\hat n,\hat l)$ to obtain
\begin{equation}
\hat q_i=\hat tq_i^t+\hat nq_i^n+\hat lq_i^l=\hat t\sin\theta_i\cos\chi_i
+\hat n\sin\theta_i\sin\chi_i+\hat l\cos\theta_i,
\end{equation}
where $\theta_i$ is the polar angle and $\chi_i$ is the azimuthal angle of the
respective charged lepton in the rest frames of the top and antitop quark.
The calculation of the trace in Eq.~(\ref{tr3rho}) results in
\begin{equation}
\rho(t)^2\left(\rho+\sum_i\rho^{P_i}q_1^{P_i}-\sum_j\rho^{P_j}q_2^{P_j}
  -\sum_{i,j}\rho^{P_iP_j}q_1^{P_i}q_2^{P_j}\right).
\end{equation}
If only the polar angles $\theta_i$ are observed, the integration over the
nonobserved azimuthal angles $\chi_i$ normalised by $(2\pi)^{-1}$ results in
\begin{equation}
\rho(t)^2\rho\left(1+\langle{\cal O}^l\rangle\cos\theta_1
  -\langle{\cal O}^l\rangle\cos\theta_2
  -\langle{\cal O}^{ll}\rangle\cos\theta_1\cos\theta_2\right),
\end{equation}
where $\langle\cdots\rangle$ denotes integration over the whole phase space.
It is apparent that one can measure the observables $\langle{\cal O}^l\rangle$
and $\langle{\cal O}^{ll}\rangle$ by analysing the polar angle dependence of
the inclusive decays $t(\uparrow)/\bar t(\uparrow)\to\ell^+/\ell^-+X$. Other
single-spin and spin--spin polarisation observables can be measured by
analysing in addition the azimuthal angular dependence of the decays
$t(\uparrow)/\bar t(\uparrow)\to \ell^+/\ell^- +X$.

The opening angle $\phi$ between the two charged leptons is defined by 
$\hat q_1\cdot\hat q_2=\cos\phi$. In order to determine the opening angle
distribution one has to integrate over all angles except for $\phi$, i.e.\ the
azimuthal angle $\chi$ of an expansion of $\hat q_2$ in terms of $\hat q_1$
and two arbitrary perpendicular directions, and over all angles that determine
the orientation of $\hat q_1$. The first normalised integration results in
\begin{eqnarray}
\lefteqn{\frac1{2\pi}\int\Tr\left(\hat\rho(t\bar t)\left(\hat\rho(t)\otimes
  \hat\rho(\bar t)\right)\right)d\chi\ =\ \frac{\rho(t)^2}{2\pi}\int\Tr
  \left(\hat\rho\left(\frac12(\oone+\hat q_1\cdot\ssigma)\otimes
  \frac12(\oone-\hat q_2\cdot\ssigma)\right)\right)d\chi}\nonumber\\&&
  \strut\kern-20pt=\frac{\rho(t)^2}4\Big\{\Tr(\hat\rho)
  +\Tr\left(\hat\rho(\hat q_1\cdot\ssigma\otimes\oone)\right)
  -\left[\Tr\left(\hat\rho(\oone\otimes\hat q_1\cdot\ssigma)\right)
  -\Tr\left(\hat\rho(\hat q_1\cdot\ssigma\otimes\hat q_1\cdot\ssigma)\right)
  \right]\cos\phi\Big\}.\nonumber\\
\end{eqnarray}
Expanding $\hat q_1=\hat t\sin\theta_1\cos\chi_1+\hat n\sin\theta_1\sin\chi_1
+\hat l\cos\theta_1$ as before, integrating over the solid angle with
$d\chi_1d(\cos\theta_1)$ and normalising by $(4\pi)^{-1}$, one finally obtains
\begin{eqnarray}
\lefteqn{\frac1{8\pi^2}\int\Tr\left(\hat\rho(t\bar t)\left(\hat\rho(t)\otimes
  \hat\rho(\bar t)\right)\right)d\chi d\chi_1d(\cos\theta_1)}\nonumber\\
  &=&\rho(t)^2\rho\left(1-\frac13\left(\langle{\cal O}^{tt}\rangle
  +\langle{\cal O}^{nn}\rangle+\langle{\cal O}^{ll}\rangle\right)\cos\phi
  \right).
\end{eqnarray}
Therefore, the trace $\Tr{\cal O}={\cal O}^{tt}+{\cal O}^{nn}+{\cal O}^{ll}$
of the three-dimensional correlation matrix in phase-space integrated form can
be determined by measuring the opening angle distribution. In
Ref.~\cite{Brandenburg:1998xw} this observable was called ${\cal O}_4$. Note
that because this observable is equally derived from the trace of $\hat\rho$
with the tensor product of the spin operator with itself, the value is equal
to $1$ at LO and decreases slightly if we include first-order radiative
corrections. Therefore, the dependence on $\cos\phi$ is at most $1/3$ of the
integrated contribution.

\section{Results up to $O(\alpha_s)$}
Before presenting our NLO results derived from the one-loop and tree-graph
contributions, we begin by presenting the Born term results. Equivalent results
have already been listed in Ref.~\cite{Brandenburg:1998xw} where, however, a
different representation has been used. The unpolarised Born term contribution
is given by
\begin{equation}
\rho=e^4N_c\Big[\left(1+v^2\cos^2\theta\right)g_{11}+(1-v^2)g_{12}
  +2v\cos\theta g_{44}\Big]
\end{equation}
($e^2=4\pi\alpha$). Replacing $g_{11}=g_{PC}^{VV}+g_{PC}^{AA}$,
$g_{12}=g_{PC}^{VV}-g_{PC}^{AA}$ and integrating over $\cos\theta$, this
result is in agreement with the well-known result 
\begin{equation}
\sigma=\frac{4\pi\alpha^2}{3q^2}N_cv\left(\frac{3-v^2}2g_{PC}^{VV}
  +v^2g_{PC}^{AA}\right).
\end{equation}
The various single-spin and spin--spin contributions have been defined in
Eqs.~(\ref{rhoee}) and~(\ref{rhoe}). They can be calculated using the
leading-order form of Eq.~(\ref{rho}). One has ($\xi=1-v^2$)
\begin{eqnarray}
\rho^t&=&-e^4N_c\sxi\sin\theta\Big[v\cos\theta g_{14}+g_{41}+g_{42}\Big],
  \nonumber\\[7pt]
\rho^n&=&-e^4N_c\sxi v\sin\theta g_{43},\nonumber\\[7pt]
\rho^l&=&e^4N_c\Big[v(1+\cos^2\theta)g_{14}
  +\cos\theta\left((1+v^2)g_{41}+(1-v^2)g_{42}\right)\Big],\nonumber\\[7pt]
\rho^{tt}&=&e^4N_c\sin^2\theta(\xi g_{11}+g_{12}),\nonumber\\[7pt]
\rho^{tn}&=&\rho^{nt}\ =\ e^4N_cv\sin^2\theta g_{13},\nonumber\\[7pt]
\rho^{tl}&=&\rho^{lt}\ =\ -e^4N_c\sxi\Big[\sin\theta\cos\theta(g_{11}+g_{12})
  +v\sin\theta g_{44}\Big],\nonumber\\[7pt]
\rho^{nn}&=&-e^4N_cv^2\sin^2\theta g_{12},\nonumber\\[7pt]
\rho^{nl}&=&\rho^{ln}\ =\ -e^4N_cv\sxi\sin\theta\cos\theta g_{13},
  \nonumber\\[7pt]
\rho^{ll}&=&e^4N_c\Big[\left(v^2+\cos^2\theta\right)g_{11}
  +\xi\cos^2\theta g_{12}+2v\cos\theta g_{44}\Big].
\end{eqnarray}
Using Eq.~(\ref{bfuvector}), the LO coefficient functions listed in
this subsection can be converted to the corresponding coefficient functions
of Ref.~\cite{Brandenburg:1998xw}.

\subsection{$O(\alpha_s)$ loop contributions}
The one-loop QCD vertex corrections have been calculated before in e.g.
Refs.~\cite{Korner:1993dy,Groote:1996nc}. They can be expressed in terms of
the two invariants $A$ and $B$ appearing in the covariant expansion of the
matrix element $\langle t\bar t|j^\mu|0\rangle$. They read 
\begin{eqnarray}
\real A&=&-\frac{\alpha_sC_F}{4\pi v}\Bigg[\left(\frac1\eps-\gamma_E
  +\ln\pfrac{4\pi\mu^2}{m^2}\right)\left(2v-(1+v^2)\ln\pfrac{1+v}{1-v}
  \right)\strut\nonumber\\&&\strut
  +(1+v^2)\left(\Li_2\pfrac{2v}{1+v}-\Li_2\pfrac{-2v}{1-v}-\pi^2\right)
  -3v^2\ln\pfrac{1+v}{1-v}+4v\Bigg],\nonumber\\
\real B&=&-\frac{\alpha_sC_F}{4\pi v}(1-v^2)\ln\pfrac{1+v}{1-v},\qquad
\imag B\ =\ \frac{\alpha_sC_F}{4\pi v}(1-v^2)\pi,\qquad
\end{eqnarray}
where $\mu$ is the renormalisation scale. The IR singularity is regularised by
the parameter $\eps=(4-D)/2$ of dimensional regularisation,
$C_F=(N_c^2-1)/(2N_c)=4/3$ for $N_c=3$, and $\gamma_E=0.577\ldots$ is the
Euler--Mascheroni constant. The dimensional IR regularisation parameter
$1/\eps$ can be converted to the gluon mass parameter used in the IR
regularisation of the tree graph integrations according to
\begin{equation}
\ln\Lambda\leftrightarrow\frac1\eps-\gamma_E+\ln\pfrac{4\pi\mu^2}{q^2},
\end{equation}
where $\Lambda=m_G^2/q^2$ is the normalised squared gluon mass. After
folding the one-loop corrections with
the Born term vertex function (spins not summed!) one obtains
\begin{eqnarray}
\rho&=&e^4N_c\Big[\left(2(1+v^2\cos^2\theta)\real A
  -v^2(1+3\cos^2\theta)\real B\right)g_{11}\strut\nonumber\\&&\strut
  +\left(2\xi\real A+v^2(3+\cos^2\theta)\real B\right)g_{12}
  \strut\nonumber\\&&\strut
  -4v\cos\theta\imag Bg_{43}+4v\cos\theta\real(A-B)g_{44}\Big],
  \nonumber\\[7pt]
\rho^{tt}&=&e^4N_c\sin^2\theta\Big[\left(2\xi\real A+3v^2\real B\right)g_{11}
  +\left(2\real A-v^2\real B\right)g_{12}\Big],\nonumber\\[7pt]
\rho^{tn}&=&\rho^{nt}\ =\ 2e^4N_cv\sin^2\theta\Big[(\real A-\real B)g_{13}
  +\imag Bg_{14}\Big],\nonumber\\[7pt]
\rho^{tl}&=&\rho^{lt}\ =\ -\frac{e^4N_c}\sxi\Big[
  \sin\theta\cos\theta\left(2\xi\real A+v^2\real B\right)(g_{11}+g_{12})\nl
  -v(1+\xi)\sin\theta\,\imag Bg_{43}+v\sin\theta(2\xi\real A+(1-3 \xi)\real B)
  g_{44}\Big],\nonumber\\[7pt]
\rho^{nn}&=&-e^4N_c v^2\sin^2\theta\Big[\real Bg_{11}
  +(2\real A-3\real B)g_{12}\Big],\nonumber\\[7pt]
\rho^{nl}&=&\rho^{ln}\ =\ -e^4N_c\frac v\sxi\Big[
  \sin\theta\cos\theta(2\xi\,\real A+(1-3\xi)\real B)g_{13}\nl
  +(1+\xi)\sin\theta\cos\theta\,\imag Bg_{14}
  -v\sin\theta\,\imag B(g_{41}+g_{42})\Big],\nonumber\\[7pt]
\rho^{ll}&=&e^4N_c\Big[\left(2\left(v^2+\cos^2\theta\right)\real A
  -v^2\left(3+\cos^2\theta\right)\real B\right)g_{11}\nl
  +\left(2\xi\cos^2\theta\,\real A
  +v^2\left(1+3\cos^2\theta\right)\real B\right)g_{12}\nl
  -4v\cos\theta\,\imag B g_{43}+4v\cos\theta(\real A-\real B)g_{44}\Big].
\end{eqnarray}

\subsection{$O(\alpha_s)$ tree-graph contributions}
According to the Lee--Nauenberg theorem, the IR singularities of the
$O(\alpha_s)$ loop calculation are canceled against the IR singularities
appearing in the tree graph calculation. In~\cite{Groote:2008ux} we
cut on the hard gluon phase space from above. In this paper we consider the
full three-particle phase space. Let us specify the kinematics of the
three-body decay more explicitly. We work in the laboratory frame with the $z$
axis defined by the top quark momentum direction. The four-momenta $q$ and
$p_1,p_2$ read
\begin{eqnarray}
q&=&\sqrt{q^2}\left(1;0,0,0\right),\nonumber\\
p_1&=&\frac12\sqrt{q^2}\left(1-y;0,0,R_y\right),\nonumber\\
p_2&=&\frac12\sqrt{q^2}\left(1-z;R_z\sin\theta_{12},0,R_z\cos\theta_{12}\right)
\end{eqnarray}
where $R_y=\sqrt{(1-y)^2-\xi}$ and $R_z=\sqrt{(1-z)^2-\xi}$.
The gluon momentum $p_3$ is given by $p_3=q-p_1-p_2$. The sine and the cosine
of the polar angle $\theta_{12}$ between the momenta of the top and antitop
quark are given by
\begin{eqnarray}
\sin\theta_{12}&=&\frac{\sqrt{4yz(1-y-z)-\xi(y+z)^2}}{R_yR_z},\nonumber\\
\cos\theta_{12}&=&-\frac{1-y-z-yz-\xi}{R_yR_z}.
\end{eqnarray}
The spin four-vectors satisfy $s_ip_i=0$ and $s_i^2=-1$. Including a sign for
the orientation, they are given by
\begin{eqnarray}
\pm s_1^L&=&\frac{\pm 1}\sxi\left(R_y;0,0,1-y\right),\nonumber\\
\pm s_2^L&=&\frac{\pm 1}\sxi\left(R_z;(1-z)\sin\theta_{12},0,
  (1-z)\cos\theta_{12}\right),\nonumber\\[3pt]
\pm s_1^T&=&\pm\left(0;1,0,0\right),\nonumber\\[7pt]
\pm s_2^T&=&\pm\left(0;\cos\theta_{12},0,-\sin\theta_{12}\right),
  \nonumber\\[7pt]
\pm s_1^N&=&\pm s_2^N\ =\ \pm\left(0;0,1,0\right).
\end{eqnarray}
Finally, the event plane spanned by the momenta of the top quark, antitop
quark and gluon is rotated with respect to the beam plane which is spanned by
the momenta of the electron and the top quark by an azimuthal angle $\chi$.
Viewed from the event plane the electron and positron have the four-momenta
\begin{equation}
p_\pm=\frac12\sqrt{q^2}\left(1;\pm\cos\chi\sin\theta,\mp\sin\chi\sin\theta,
  \mp\cos\theta\right).
\end{equation}
We do not list the explicit forms of the tree-graph contributions but merely
catalog the generic structure of the integrals that appear in the 
phase space integration. The basic integrals have the structure
\begin{equation}
I_{n_yn_z}(m_y,m_z)=\int_{y_-}^{y_+}dy\int_{z_-(y)}^{z_+(y)}dz\,
y^{m_y}z^{m_z}R_y^{n_y}R_z^{n_z},
\end{equation}
where $n_y$ ranges from $0$ to $-4$ and $n_z$ takes the values $0$ and $-2$.
The indices $m_y,m_z$ are limited from the below by $m_y+m_z\ge -2$
and $m_y,m_z\ge -2$. The integrals with $m_y+m_z=-2$ are IR singular. For the
regularisation of the IR
singularity at $y=z=0$ we use a finite gluon mass $m_G=\sqrt{\Lambda q^2}$,
such that the phase space limits are now given by
$y_-=\Lambda+\sqrt{\Lambda\xi}$, $y_+=1-\sxi$ and
\begin{equation}
z_\pm(y)=\frac1{4y+\xi}\left(2y-2y^2-\xi y+2\Lambda y+2\Lambda
  \pm 2R_y\sqrt{(y-\Lambda)^2-\Lambda\xi}\right).
\end{equation}
The subtraction of the singularity is performed by adding and subtracting an
integral with the same singular behaviour but with a simpler integrand. The
simplified integrand is obtained from the original integrand by an expansion
around $y=0$. In this expansion, both $R_y$ and $R_z$ are
replaced by
$v$, leading to the generic divergent parts
\begin{eqnarray}\label{gendiv}
I_D(-2,0)\!\!\!&=&\!\!\!\frac{4v}\xi\int_{\Lambda+\sqrt{\Lambda\xi}}^{1-\sxi}
  \frac{dy}{y^2}\sqrt{(y-\Lambda)^2-\Lambda\xi}
  =\frac{4v}\xi\left(\ln\pfrac{2(1-\sxi)}{\sqrt{\Lambda\xi}}-1\right),
  \nonumber\\
I_D(0,-2)\!\!\!&=&\!\!\!\int_{\Lambda+\sqrt{\Lambda\xi}}^{1-\sxi}dy
  \left(\frac1{z_-^s(y)}-\frac1{z_+^s(y)}\right)
  =\frac{4v}\xi\ln\pfrac{2(1-\sxi)}{\sqrt{\Lambda\xi}}
  -2\frac{1+v^2}{1-v^2}\ln\pfrac{1+v}{1-v},\nonumber\\
I_D(-1,-1)\!\!\!&=&\!\!\!\int_{\Lambda+\sqrt{\Lambda\xi}}^{1-\sxi}\frac{dy}y
  \ln\pfrac{z_+^s(y)}{z_-^s(y)}=I_D-\ln\pfrac{1+v}{1-v}\ln\Lambda,
\end{eqnarray}
where $\xi z_\pm^s(y)=(1+v^2)y\pm 2v\sqrt{(y-\Lambda)^2-\Lambda\xi}$ and
\begin{eqnarray}
I_D&:=&2\ln\pfrac{1-\sxi}\sxi\ln\pfrac{1+v}{1-v}
  -\Li_2\pfrac{2v}{(1+v)^2}+\Li_2\pfrac{-2v}{(1-v)^2}\strut\nonumber\\&&\strut
  +\frac12\Li_2\left(-\frac{(1-v)^2}{(1+v)^2}\right)
  -\frac12\Li_2\left(-\frac{(1+v)^2}{(1-v)^2}\right).
\end{eqnarray}
The results for the generic divergent parts in Eqs.~(\ref{gendiv}) are only
accurate up to power-suppressed terms in $\sqrt\Lambda$. Adding and
subtracting these divergent parts (including a corresponding factor
$v^{n_y+n_z}$), the difference between the original integral and the divergent
part turns out to be IR finite. This is equivalent to adding 
counterterms to the original unregularised integrals. The counterterms take
the form 
\begin{eqnarray}
I_D(-2,0),I_D(0,-2)&\to&\frac{4v}\xi\int_0^{1-\sxi}\frac{dy}y,\nonumber\\
I_D(-1,-1)&\to&2\ln\pfrac{1+v}{1-v}\int_0^{1-\sxi}\frac{dy}y.
\end{eqnarray}
After having removed the IR singularities it is not difficult to do the
$z$ integration. One encounters integrals of the form
\begin{equation}
\int z^{-1}R_z^{-2}dz=\frac{\ln z}{1-\xi}+\frac{\ln(1-z+\sxi)}{2(1+\sxi)\sxi}
  -\frac{\ln(1-z-\sxi)}{2(1-\sxi)\sxi}.
\end{equation}
In the subsequent $y$ integration one encounters integrands of the form
\begin{equation}
L_0=\Big[\ln z\Big]_{z_-(y)}^{z_+(y)},\quad
L_\pm=\Big[\ln\left(1-z\pm\sxi\right)\Big]_{z_-(y)}^{z_+(y)}.
\end{equation}
To do the $y$ integration one uses the substitution
\begin{equation}\label{subst}
y=1-\frac\sxi2\left(t+\frac1t\right)
\end{equation}
which allows one to factorise the arguments of the logarithms, e.g.
\begin{eqnarray}\label{LRdef}
dy&=&-\frac\sxi2\left(1-\frac1{t^2}\right)\,dt,\qquad
R_y\ =\ \frac\sxi2\left(\frac1t-t\right),\nonumber\\
L_0(t)&=&\ln\pfrac{2-\sxi t}{t(2t-\sxi)},\nonumber\\
L_+(t)&=&\ 2\ln\pfrac{(2+\sxi)t-\sxi}{2+\sxi-\sxi t}
  +\ln\pfrac{2-\sxi t}{t(2t-\sxi)},\nonumber\\
L_-(t)&=&\ 2\ln\pfrac{|(2-\sxi)t-\sxi|}{2-\sxi-\sxi t}
  +\ln\pfrac{2-\sxi t}{t(2t-\sxi)}.
\end{eqnarray}
Note that for integrals containing $L_-$ the substitution needs a subdivision
of the integration interval $[t_-,t_1]$ into two parts, divided by the point
$t_0$ where
\begin{equation}
t_\pm=\frac{1\pm v}\sxi,\qquad
t_0=\frac\sxi{2-\sxi},\qquad
t_+t_-=1.
\end{equation}
It came as a surprise to us that in the end the contributions
containing $L_-$ in the integrand cancel (for a discussion, see
Appendix~A). One is finally left with three types of integrals
\begin{eqnarray}\label{Cintdef}
I^L_0(n)&:=&\int L(t)t^{n-1}dt,\\
I^L_{1\pm}(n)&:=&\int L(t)(1\pm t)^{n-1}dt,\\
I^L_{t\pm}(n)&:=&\int L(t)t_{\pm}^{-n}(t-t_\pm)^{n-1}dt,
\end{eqnarray}
where $L$ stands for $L_0$, $L_+$ or can be skipped if no logarithm appears
in the integrand. Our final results contain only those integrals (with
$L=L_0,L_+$ and $n=0$) which contain dilogarithms. All other contributions are
at most logarithmic. These dilogarithmic integrals, together with the standard
logarithmic integrals, are found in Appendix~A.

\subsection{$O(\alpha_s)$ total contributions}
When one adds the $O(\alpha_s)$ loop and tree-graph contributions, one obtains
IR-finite results. The list of results is quite long and they are presented in
Appendix~B. As an illustrative example we list the NLO unpolarised rate 
$\rho=g_{11}\rho_{11}+g_{12}\rho_{12}+g_{43}\rho_{43}+g_{44}\rho_{44}$ where
the coefficients $\rho_{ij}$ are given by
\begin{eqnarray}
\rho_{11}&=&N\Bigg[
  \frac v4\left((10-\xi)(2-3\xi)-3(4-20\xi+\xi^2)\cos^2\theta\right)
  \strut\nonumber\\&&\strut
  +\frac18\left(96-8\xi-18\xi^2+3\xi^3+(6-\xi)(16-28\xi+3\xi^2)\cos^2\theta
  \right)\ell_3\strut\nonumber\\&&\strut
  -\frac12\left(8-2\xi+3\xi^2+(4+\xi)(2-5\xi)\cos^2\theta\right)I^{L_0}_0(0)
  \strut\nonumber\\&&\strut
  -\frac\sxi2(1-\sxi)(2+4\sxi-3\xi)(1-3\cos^2\theta)I^{L_0}_{1-}(0)
  \strut\nonumber\\&&\strut
  -\frac\sxi2(1+\sxi)(2-4\sxi-3\xi)(1-3\cos^2\theta)I^{L_0}_{1+}(0)
  \strut\nonumber\\&&\strut
  +2(1+v^2\cos^2\theta)\left((1+v^2)(\hat I^{L_0}_{t-}(0)+I^{L_0}_{t+}(0))
  -4v(\ell_0^-+\ell_0^+)\right)\Bigg],\\[7pt]
\rho_{12}&=&N\xi\Bigg[\frac{3v}4\left(14-\xi-(6-\xi)\cos^2\theta\right)
  \strut\nonumber\\&&\strut
  +\frac18\left(48-20\xi-3\xi^2+3\xi(4+\xi)\cos^2\theta\right)\ell_3
  -\frac12\left(8-5\xi+3\xi\cos^2\theta\right)I^{L_0}_0(0)
  \strut\nonumber\\&&\strut
  -\frac\sxi2(1-\sxi)(1-3\cos^2\theta)I^{L_0}_{1-}(0)
  -\frac\sxi2(1+\sxi)(1-3\cos^2\theta)I^{L_0}_{1+}(0)\strut\nonumber\\&&\strut
  +2\left((1+v^2)(\hat I^{L_0}_{t-}(0)+I^{L_0}_{t+}(0))
  -4v(\ell_0^-+\ell_0^+)\right)\Bigg],\\[7pt]
\rho_{43}&=&N\xi\Big[-4\pi v\cos\theta\Big],\\[7pt]
\rho_{44}&=&N\Bigg[
  -8\sxi(1-\sxi)+16\ell_2+4v(2-3\xi)\ell_3\strut\nonumber\\&&\strut
  -2(4-5\xi)I^{L_0}_0(0)+4v\left((1+v^2)(\hat I^{L_0}_{t-}(0)-I^{L_0}_{t+}(0))
  -4v\ell_0^-\right)\Bigg]\cos\theta.
\end{eqnarray}
The common factor $N$ is given by\begin{equation}\label{genfac}
N=e^4N_c\frac{\alpha_sC_F}{4\pi v}
\end{equation}
which is composed of the overall Born term factor $e^4N_c$, the strong
coupling factor $\alpha_s$, the color factor
$C_F$ and the relative
three-/two-particle phase space factor $(4\pi v)^{-1}$. 

\section{Numerical results and comparison}
In order to check on our analytical results we compare them with the numerical
results of Refs.~\cite{Brandenburg:1998xw,Brandenburg:1999ss}. Using the same
values for the parameters as given in Refs.~\cite{Brandenburg:1998xw,%
Brandenburg:1999ss}, we reproduce the entries of Table~I in
Ref.~\cite{Brandenburg:1998xw} with an accuracy of $0.2\%$. We also
agree on the various figures presented in Ref.~\cite{Brandenburg:1999ss}.
We remind the reader that we employ an orthonormal frame
$(\hat t,\hat n,\hat l)$ instead of the nonorthogonal frame
$(\hat k,\hat p,\hat n)$ of Ref.~\cite{Brandenburg:1998xw} and the
un-normalised but orthogonal frame $(\hat k,\hat k^\perp,\hat n)$ of
Ref.~\cite{Brandenburg:1999ss}.

\subsection{Polar angle dependence for different energies}
\begin{figure}\begin{center}
\epsfig{figure=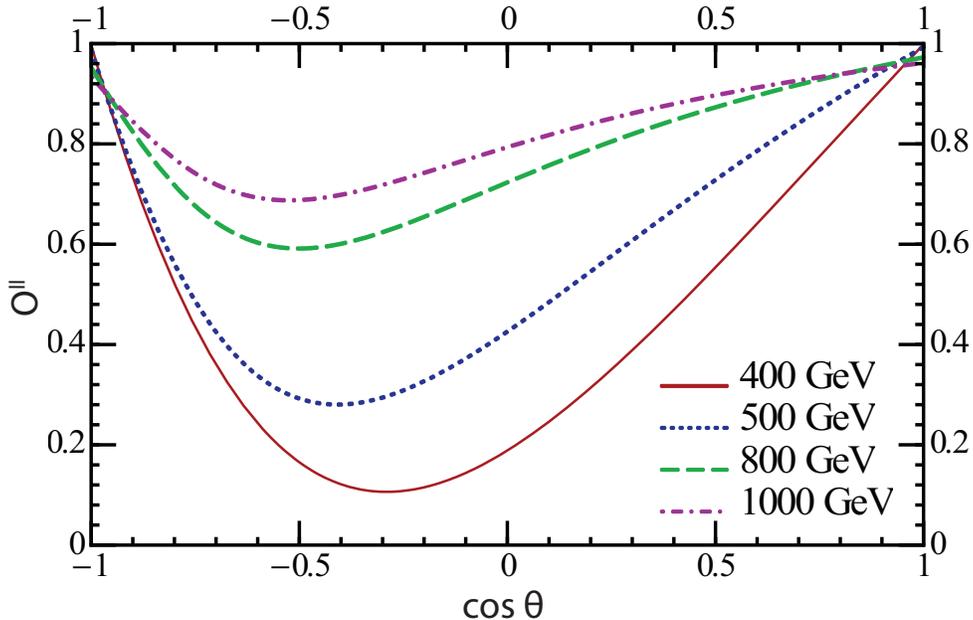, scale=0.8}
\end{center}
\caption{\label{figOll} The observable ${\cal O}^{ll}$ as a function of
$\cos\theta$ for different energies $\sqrt{q^2}=400$ (solid line), $500$
(dotted), $800$ (dashed), and $1000\GeV$ (dashed dotted)}
\end{figure}

\begin{figure}\begin{center}
\epsfig{figure=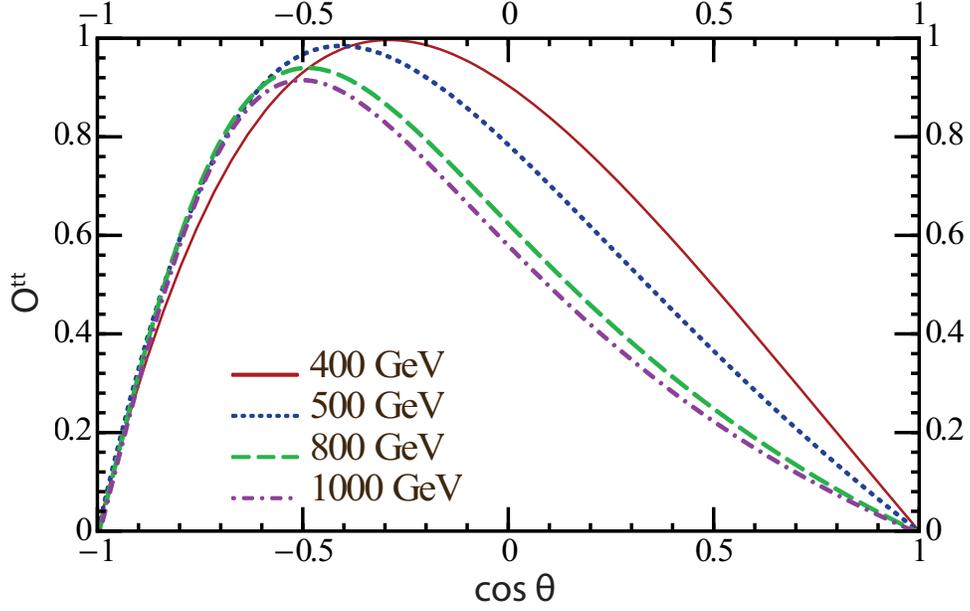, scale=0.8}
\end{center}
\caption{\label{figOtt}The observable ${\cal O}^{tt}$ as a function of
$\cos\theta$ for different energies $\sqrt{q^2}=400$ (solid line), $500$
(dotted), $800$ (dashed), and $1000\GeV$ (dashed dotted)}
\end{figure}

\begin{figure}\begin{center}
\epsfig{figure=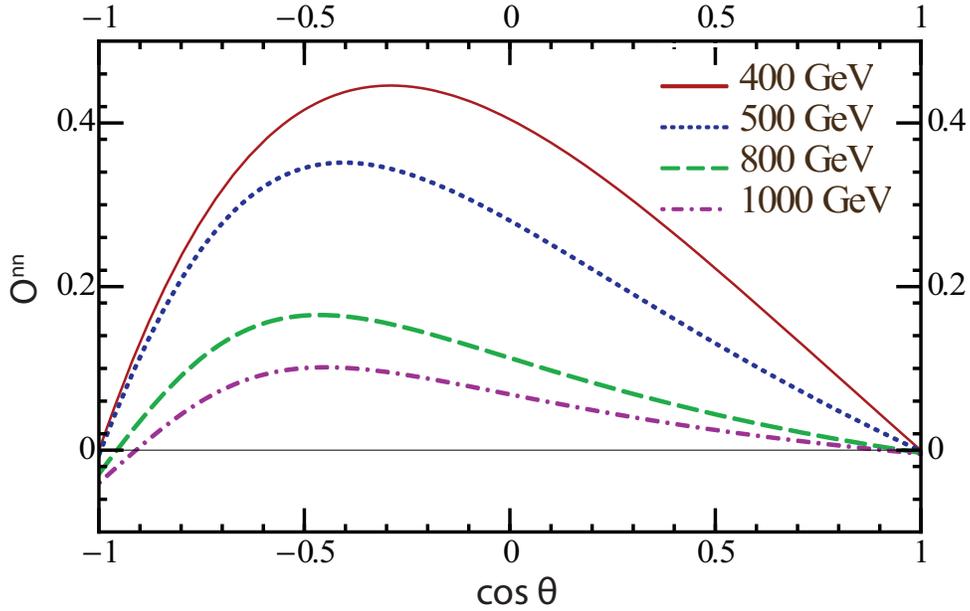, scale=0.8}
\end{center}
\caption{\label{figOnn}The observable ${\cal O}^{nn}$ as a function of
$\cos\theta$ for different energies $\sqrt{q^2}=400\GeV$ (solid line),
$500\GeV$ (dotted), $800\GeV$ (dashed), and $1000\GeV$ (dashed dotted)}
\end{figure}

\begin{figure}\begin{center}
\epsfig{figure=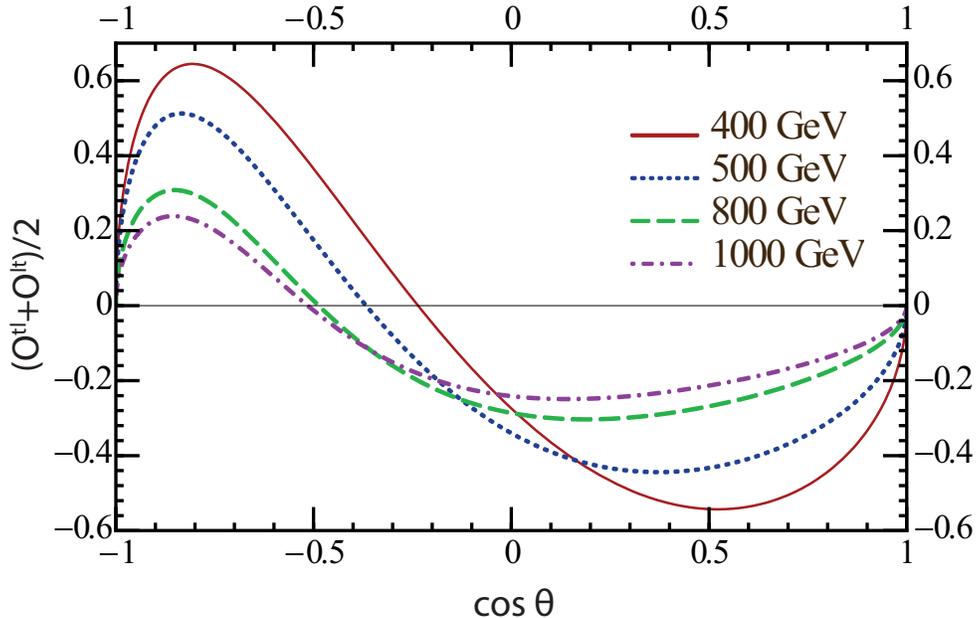, scale=0.8}
\end{center}
\caption{\label{figOlt}$({\cal O}^{tl}+{\cal O}^{lt})/2$ as a function of
$\cos\theta$ for different energies $\sqrt{q^2}=400$ (solid line), $500$
(dotted), $800$ (dashed), and $1000\GeV$ (dashed dotted)}
\end{figure}

\begin{figure}\begin{center}
\epsfig{figure=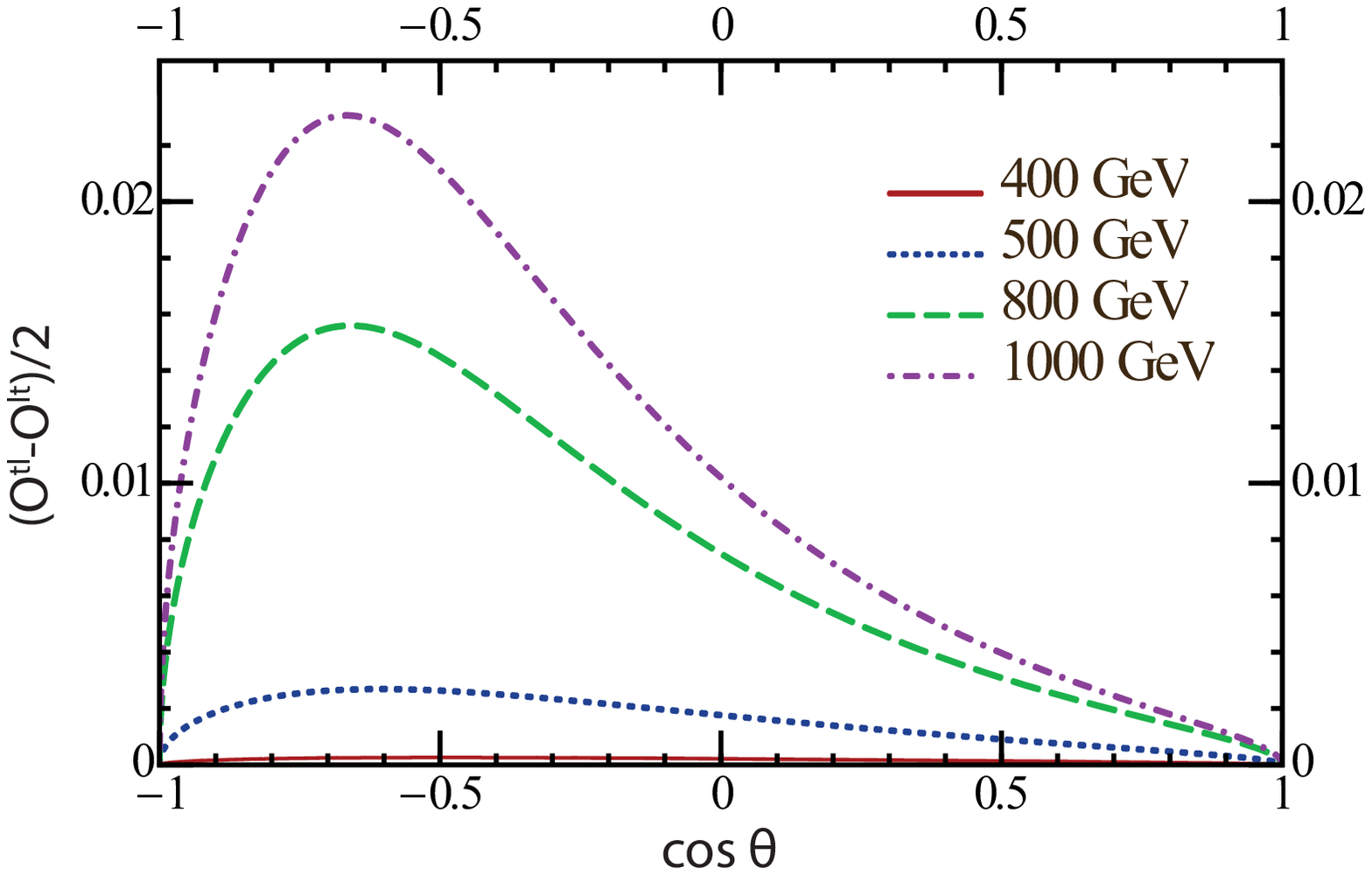, scale=0.75}
\end{center}
\caption{\label{figOtl}$({\cal O}^{tl}-{\cal O}^{lt})/2$ as a function of
$\cos\theta$ for different energies $\sqrt{q^2}=400\GeV$ (solid line),
$500\GeV$ (dotted), $800\GeV$ (dashed), and $1000\GeV$ (dashed dotted)}
\end{figure}

\begin{figure}\begin{center}
\epsfig{figure=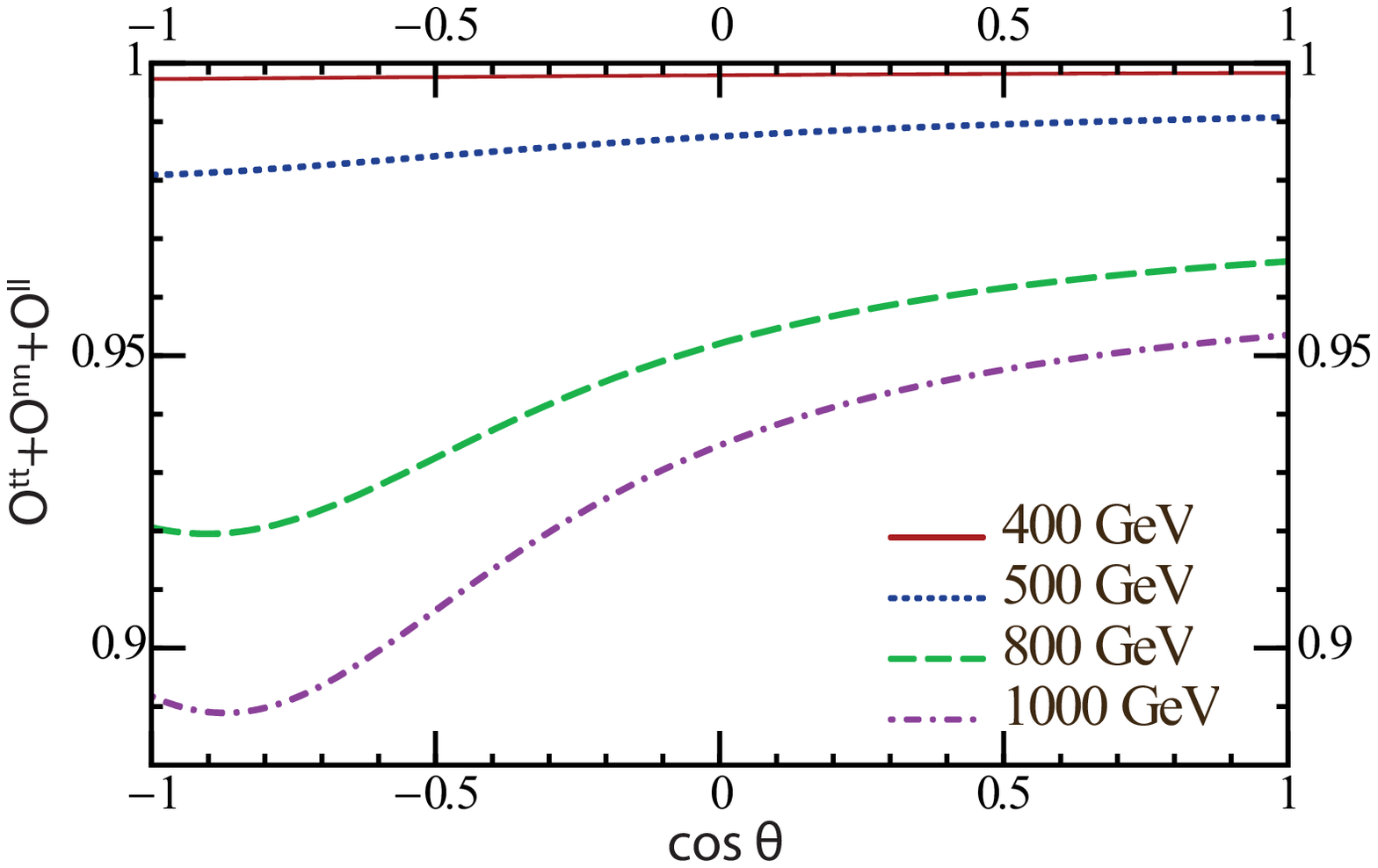, scale=0.75}
\end{center}
\caption{\label{figOtra}Trace $\Tr{\cal O}$ as a function of $\cos\theta$ for
different energies $\sqrt{q^2}=400$ (solid line), $500$ (dotted), $800$
(dashed), and $1000\GeV$ (dashed dotted)}
\end{figure}

In Figs.~\ref{figOll}--\ref{figOtra} we present our results for some of the
observables defined in Eq.~(\ref{defobs}). The full set of nine
observables for the correlation matrix can be divided up into the diagonal
elements ${\cal O}^{tt}$, ${\cal O}^{nn}$ and ${\cal O}^{ll}$, and a set of
six off-diagonal elements. For the parameters we use the
values~\cite{Agashe:2014kda}
\begin{eqnarray}
m_Z=91.1876(21)\GeV,&\Gamma_Z=2.4952(23)\GeV,&m_t=174.6(1.9)\GeV\\[7pt]
\sin^2\theta_W=0.23126(5),&\alpha_S(m_Z)=0.1185(6),&
G_F/(\hbar c)^3=1.1663787(6)\times 10^{-5}\GeV^{-2}.\nonumber
\end{eqnarray}

In Fig.~\ref{figOll} we plot the polar angle dependence of the dominant
diagonal element ${\cal O}^{ll}$. The value is close to $100\%$ in the forward
and backward directions and decreases slightly with increasing center-of-mass
energy $\sqrt{q^2}$. In the transverse direction the value falls off to $10\%$
for $\sqrt{q^2}=400\GeV$ and only to $70\%$ for $\sqrt{q^2}=1000\GeV$ while the
location of the minimum tends to the backward direction.

While the depth of the minimum for ${\cal O}^{ll}$ decreases with increasing
c.m.\ energy $\sqrt{q^2}$, the situation is reversed for ${\cal O}^{tt}$
shown in Fig.~\ref{figOtt}. In the forward and backward directions the value
is exactly zero, as there is no boost that can turn the transversal direction
into the direction of the quark, while the maximal value slightly falls from
nearly $100\%$ for $\sqrt{q^2}=400\GeV$ to $90\%$ for $\sqrt{q^2}=1000\GeV$.
The position of the maximum of ${\cal O}^{tt}$ roughly coincides with the
position of the minimum of ${\cal O}^{ll}$. 

The remaining diagonal element ${\cal O}^{nn}$ is again exactly zero in the
forward and backward directions. As shown in Fig.~\ref{figOnn}, the maximum of
the absolute value of this (negative) observable ${\cal O}^{nn}$ increases
from $10\%$ for $\sqrt{q^2}=400\GeV$ to $70\%$ for $\sqrt{q^2}=1000\GeV$,
with nearly the same position of the extremum as for ${\cal O}^{ll}$ and
${\cal O}^{tt}$.

Looking at the off-diagonal elements, the values for the observables
${\cal O}^{tl}$ and ${\cal O}^{lt}$ are very close, albeit not equal. In
Fig.~\ref{figOlt} the mean value $({\cal O}^{tl}+{\cal O}^{lt})/2$ of
these two observables is displayed. The value again vanishes in the forward
and backward directions. The sine-type run of the curve for low energies is
again shifted to the backward direction for higher energies while the absolute
values are falling. The position of the zero crossing coincides again roughly
with the positions of the extrema in the previous diagrams.

In Fig.~\ref{figOtl} we show the normalised difference of the two adjoint
nondiagonal elements, $({\cal O}^{tl}-{\cal O}^{lt})/2$. Obviously, the
normalised difference vanishes for small energies and increases to a maximum
value of $2.3\%$ for $\sqrt{q^2}=1000\GeV$. However, for increasing energies
the position of the maximum stays at a nearly constant value of approximately
$\cos\theta=-0.67$.

In Fig.~\ref{figOtra} the angular dependence of the trace
$\Tr{\cal O}={\cal O}^{tt}+{\cal O}^{nn}+{\cal O}^{ll}$ of the
three-dimensional correlation matrix relevant for the opening angle
distribution is shown. At threshold (and also at LO) the trace is $1$ while
for higher energies the value decreases especially in the backward direction.

\subsection{Dependence on initial beam polarisation}
\begin{figure}\begin{center}
\epsfig{figure=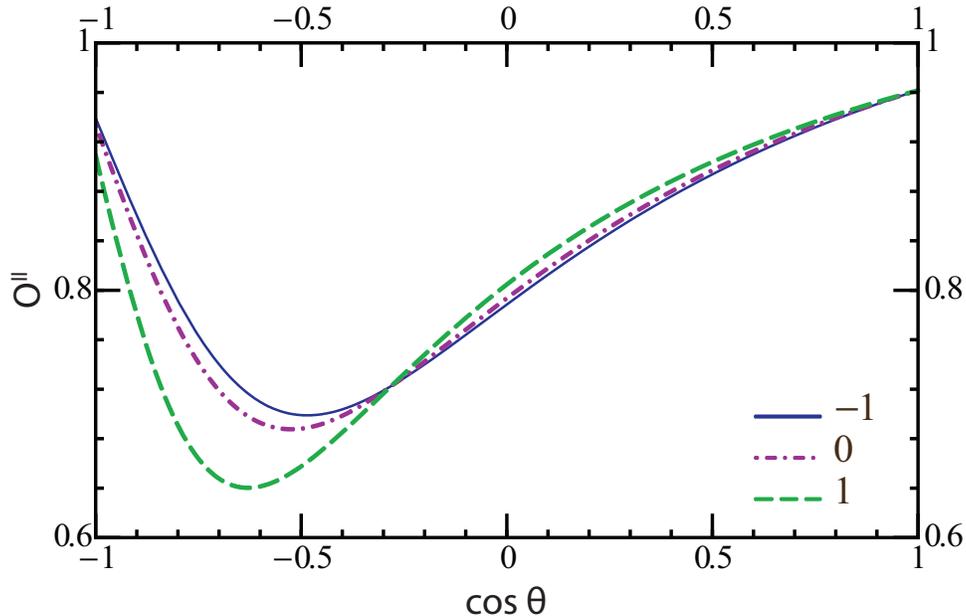, scale=0.8}
\end{center}
\caption{\label{figOllh}The observable ${\cal O}^{ll}$ as a function of
$\cos\theta$ for $\sqrt{q^2}=1000\GeV$ and different initial beam
polarisations, given by $P_{\rm eff}=0,\pm 1$}
\end{figure}

The initial beam polarisation can be easily implemented by changing the
electroweak coupling factors $g_{ij}$ according to
\begin{eqnarray}
g_{1j}\to(1-h_-h_+)g_{1j}+(h_--h_+)g_{4j},\qquad
g_{4j}\to(h_--h_+)g_{1j}+(1-h_-h_+)g_{4j},
\end{eqnarray}
where $h_-$ and $h_+$ are twice the helicities of the initial electron and
positron beams, respectively~\cite{Groote:2010zf}. For the normalised density
matrix elements one remains with the single parameter dependence given
by~Ref.~\cite{Groote:2010zf}
\begin{equation}
P_{\rm eff}=\frac{h_--h_+}{1-h_-h_+}.
\end{equation}
It is clear that the two limiting cases $P_{\rm eff}=\pm 1$ cannot be
realised technically. However, since the polarisation effects are governed by
$P_{\rm eff}$ and {\it not} by $h_\mp=-h_\pm=\mp1$, one can get very close to
the limiting cases $P_{\rm eff}=\mp 1$ with presently achievable degrees of
beam polarisation of $\sim 80\,\%$ (see the discussion in
Ref.~\cite{Groote:2010zf}). For example, for $h_-=-h_+=-0.8$ one has
$P_{\rm eff}=-0.976$. In Fig.~\ref{figOllh} we show the dependence of
${\cal O}^{ll}$ on the polar angle for $\sqrt{q^2}=1000\GeV$ and the values
$P_{\rm eff}=0,\pm 1$. The dependence of the spin-spin correlation on the
initial beam polarisation turns out to be much smaller than the dependence of
the single-spin polarisation (cf.\ Ref.~\cite{Groote:2010zf}). For
${\cal O}^{ll}$ it amounts to $10\%$ close to $\cos\theta=-0.7$ while the
single-spin observable can change locally by more than $100\%$.

\subsection{Comparison with other publications}
As mentioned earlier, we have checked that the single-spin observables are
analytically identical to the one presented in Refs.~\cite{Groote:1995yc,%
Groote:1996nc,Groote:1995ky}. This by-product in the present investigation
provides a profound cross-check for our calculation. Because of the use of
different observables, the results of Refs.~\cite{Tung:1997ur,Groote:1997su,%
Groote:2009zk} cannot be directly compared with ${\cal O}^{ll}$.\footnote{In
Refs.~\cite{Tung:1997ur,Groote:1997su,Groote:2009zk}, $\rho^{L\bar L}$ has
been investigated.} On the other hand, we obtain good agreement
with the diagrams presented in Ref.~\cite{Brandenburg:1999ss}.

\section{Results in the bases of Parke and Shadmi}
Parke and Shadmi have discussed the case of completely polarised beams which
leads to very simple rate and polarisation formulas~\cite{Parke:1996pr}. They
considered the two cases $LR$ ($h_-=-1$, $h_+=1$) and $RL$ ($h_-=1$, $h_+=-1$).
Here we concentrate on the case $LR$. Parke and Shadmi have introduced the
chiral electroweak coupling factors $f_{LL}$ and $f_{LR}$ which can be related
to our coupling factors $g_{ij}$ by~\cite{Groote:2010zf}
\begin{eqnarray}\label{fLLLR}
  f_{LL}&=&\sqrt{g_{11}-g_{14}-g_{41}+g_{44}}
  =Q_eQ_f+|\chi_Z|(v_e+a_e)(v_f+a_f),\nonumber\\
  f_{LR}&=&\sqrt{g_{11}+g_{14}-g_{41}-g_{44}}
  =Q_eQ_f+|\chi_Z|(v_e+a_e)(v_f-a_f).
\end{eqnarray}
Since the production threshold for top quarks of mass $175\GeV$ is far above
the $Z$ boson pole, we neglect the $Z$ width and we have therefore dropped the
contributions of the coupling factors $g_{13,23}$, $g_{31,32}$ and $g_{34,43}$.
For the case $LR$ one has the following replacements:
\begin{eqnarray}\label{beampol}
  g_{11}&\to&2g_{11}-2g_{41}=f^2_{LL}+f^2_{LR},\nonumber\\ 
  g_{12}&\to&2g_{12}-2g_{42}=2f_{LL}f_{LR},\nonumber\\
  g_{14}&\to&2g_{14}-2g_{44}=-(f^2_{LL}-f^2_{LR}),\nonumber\\
  g_{41}&\to&2g_{41}-2g_{11}=-(f^2_{LL}+f^2_{LR}),\nonumber\\
  g_{42}&\to&2g_{42}-2g_{12}=-2f_{LL}f_{LR},\nonumber\\
  g_{44}&\to&2g_{44}-2g_{14}=f^2_{LL}-f^2_{LR}.
\end{eqnarray}
For some frequently occurring linear combinations one obtains 
\begin{eqnarray}
  g_{11}+g_{12}&\to&+(f_{LL}+f_{LR})^2,\nonumber\\ 
  g_{11}-g_{12}&\to&+(f_{LL}-f_{LR})^2,\nonumber\\
  g_{41}+g_{42}&\to&-(f_{LL}+f_{LR})^2,\nonumber\\
  g_{41}-g_{42}&\to&-(f_{LL}-f_{LR})^2.
\end{eqnarray}
Using the abbreviations ($v=\sqrt{1-4m^2/q^2}$)
\begin{eqnarray}\label{atl}
A_{LR}&=&f_{LL}(1+v\cos\theta)+f_{LR}(1-v\cos\theta),\nonumber\\ 
T&=&\sin\theta\sqrt{1-v^2}(f_{LL}+f_{LR}),\nonumber\\
L&=&f_{LL}(\cos\theta+v)+f_{LR}(\cos\theta-v),
\end{eqnarray}
where $A_{LR}^2=T^2+L^2+4f_{LL}f_{LR}v^2\sin^2\theta$ and
\begin{equation}\label{relLR}
(1\pm\cos\theta)^2\left(f_{LL}(1\pm v)+f_{LR}(1\mp v)\right)^2=(A_{RL}\pm L)^2,
\end{equation}
we shall present our Born term results on the single- and double-spin density
matrices in two different coordinate systems. These are the helicity basis and
the off-diagonal basis introduced in Ref.~\cite{Parke:1996pr}.

\subsection{Helicity basis}
Rewriting the Born term results of Sec.~3 in terms of the chiral coupling
factors $f_{LL}$ and $f_{LR}$, the nonvanishing contributions are given by
\begin{eqnarray}
\rho&=&+e^4N_c\,(T^2+L^2+2f_{LL}f_{LR}\,v^2\sin^2\theta),\nonumber\\
\rho^t&=&+e^4N_c\,T\,A_{LR},\nonumber\\
\rho^l&=&-e^4N_c\,L\,A_{LR},\nonumber\\ 
\rho^{tt}&=&+e^4N_c\,(T^2+2f_{LL}f_{LR}\,v^2\sin^2\theta),\nonumber\\
\rho^{tl}&=&\rho^{lt}\ =\ -e^4N_c\, LT,\nonumber\\
\rho^{nn}&=&-2e^4N_c\,v^2\sin^2\theta\,f_{LL}f_{LR},\nonumber\\
\rho^{ll}&=&+e^4N_c\,(L^2+2f_{LL}f_{LR}\,v^2\sin^2\theta).
\end{eqnarray}
From the single-spin density matrix elements $\rho^t$ and $\rho^l$ one can
calculate the angle between the polarisation vector of the top quark and the
direction of the top quark. For definiteness we call this angle $\theta_{LR}$.
One has
\begin{equation}\label{offdiag}
  \frac{\sin\theta_{LR}}{\cos\theta_{LR}}=\frac{\rho^t}{\rho^l}
  =-\frac{T\,A_{LR}}{L\,A_{LR}}=-\frac{T}{L}.
\end{equation}
We shall see that the direction of the polarisation vector of the top quark
defines the $z$ direction of the off-diagonal basis of Parke and Shadmi.

In order to check on our results we calculate the rates
$\rho(\uparrow,\uparrow)$, $\rho(\uparrow,\downarrow)$,
$\rho(\downarrow,\uparrow)$ and $\rho(\downarrow,\downarrow)$. Note that
Parke and Shadmi have defined the quantisation axes as themomentum axes of the
top and antitop quark, respectively. Therefore, the second arrow has to be
reinterpreted to fit with our convention. Using both conventions, at the Born
term level one obtains
\begin{equation}\label{tutu}
\rho(t_\uparrow\bar t_\uparrow)\ =\ \rho(\uparrow,\downarrow)
  \ =\ \frac{1}{4}(\rho+\rho^{l_1}-\rho^{l_2}
  -\rho^{l_1l_2})\ =\ \frac {1}{4}(\rho-\rho^{ll})\ =\ \frac14e^4N_c\,T^2,
\end{equation}
and $\rho(t_\downarrow\bar t_\downarrow)=\rho(\downarrow,\uparrow)
=\rho(\uparrow,\downarrow)=\rho(t_\uparrow\bar t_\uparrow)$ since
$\rho^{\ell_1}=\rho^{\ell_2}=\rho^l$. Furthermore, one obtains
\begin{eqnarray}
\rho(t_\uparrow\bar t_\downarrow)\ =\ \rho(\uparrow,\uparrow)
  &=&\frac14(\rho+\rho^{l_1}+\rho^{l_2}
  +\rho^{l_1l_2})\ =\ \frac14(\rho+2\rho^l+\rho^{ll})\nonumber\\
  &=&\frac14e^4N_c\,\Big((1-\cos\theta)^2(f_{LL}(1-v)+f_{LR}(1+v))^2\Big),
  \nonumber\\
\rho(t_\downarrow\bar t_\uparrow)\ =\ \rho(\downarrow,\downarrow)
  &=&\frac{1}{4}(\rho-\rho^{l_1}-\rho^{l_2}
  +\rho^{l_1l_2})\ =\ \frac14(\rho-2\rho^l+\rho^{ll})\nonumber\\
  &=&\frac14e^4N_c\,\Big((1+\cos\theta)^2
  \left(f_{LL}(1+v)+f_{LR}(1-v)\right)^2\Big).
\end{eqnarray}
where we have used the relation~(\ref{relLR}).
All rates agree with the results of Parke and Shadmi.

\subsection{Off-diagonal basis}
Parke and Shadmi have introduced an off-diagonal basis by demanding
that the rates $\rho(t_\uparrow\bar t_\uparrow)$ and
$\rho(t_\downarrow\bar t_\downarrow)$ vanish in that basis. We show that
this demand leads to the condition that the angle $\theta_{LR}$ between the
helicity basis and the off-diagonal basis is determined by
Eq.~(\ref{offdiag}). Let us consider the rate
$\rho(t_\uparrow\bar t_\uparrow)'$ in the off-diagonal basis\footnote{Note
that the direction of the active angle $\theta_{LR}$ is opposed to the
direction of the passive angle $\theta$.}
\begin{equation}
\vec e^{\,l\,'}=\cos\theta_{LR}\vec e^{\,l}+\sin\theta_{LR}\vec e^{\,t},\qquad
\vec e^{\,t\,'}=-\sin\theta_{LR}\vec e^{\,l}+\cos\theta_{LR}\vec e^{\,t}.
\end{equation}
Using the spin projection formula~(\ref{rhoee}) and Eq.~(\ref{tutu}), at the
Born term level one has
\begin{eqnarray}
\rho(t_\uparrow\bar t_\uparrow)'
  &=&\frac14\big(\rho-\cos^2\theta_{LR}\rho^{ll}-\sin^2\theta_{LR}\rho^{tt}
  -2\sin\theta_{LR}\cos\theta_{LR}\rho^{lt}\big)\nonumber\\
  &=&\frac14e^4N_c(L\sin\theta_{LR}+T\cos\theta_{LR})^2.
\end{eqnarray}
Therefore, the condition $\rho(t_\uparrow\bar t_\uparrow)'=0$ defines the
off-diagonal basis via the condition $L\sin\theta_{LR}+T\cos\theta_{LR}=0$.

\subsection{$O(\alpha_s)$ corrections}
Parke and Shadmi~\cite{Parke:1996pr} expressed their expectations that
radiative corrections to their analysis of spin-spin correlations in
$e^+e^-$ annihilations are small. An attempt to estimate the effects of
radiative corrections were published in Ref.~\cite{Kodaira:1998gt}. However,
since the calculation of Ref.~\cite{Kodaira:1998gt} employed the soft-gluon
approximation, the results are incomplete in the sense that spin-spin
correlation effects due to hard gluon emission are missed in such a
calculation. The present paper rectifies this omission.

Going beyond the Born term level, the simultaneous disappearance of
$\rho^{t\,'}$ and $\rho(t_\uparrow\bar t_\uparrow)'$ is no longer granted.
In this case one has to decide which of the two quantities is used to define
the angle $\theta_{LR}$. Here we decide to define the angle via the
single-spin quantity, as it was done in Ref.~\cite{Groote:2010zf}. This
definition is physically more transparent. Furthermore it is also valid for
unpolarised or partly polarised beams. In addition to the polar angle
$\theta_{LR}$, an azimuthal angle $\chi_{LR}$ has to be defined which points
out of the plane spanned by the electron and top quark momenta. Actually, this
azimuthal angle is already present at the Born term level if one takes into
account the coupling factors $g_{13}$ and $g_{43}$. The angles can be
unambiguously defined by
\begin{equation}
\sin\theta_{LR}=\frac{\rho^t}{\sqrt{(\rho^t)^2+(\rho^l)^2}},\qquad
\cos\theta_{LR}=\frac{\rho^l}{\sqrt{(\rho^t)^2+(\rho^l)^2}}
\end{equation}
and
\begin{equation}
\sin\chi_{LR}=\frac{\rho^n}{\sqrt{(\rho^t)^2+(\rho^n)^2+(\rho^l)^2}},\qquad
\cos\chi_{LR}=\frac{\sqrt{(\rho^t)^2+(\rho^l)^2}}{\sqrt{(\rho^t)^2+(\rho^n)^2
  +(\rho^l)^2}}
\end{equation}
\begin{figure}\begin{center}
\epsfig{figure=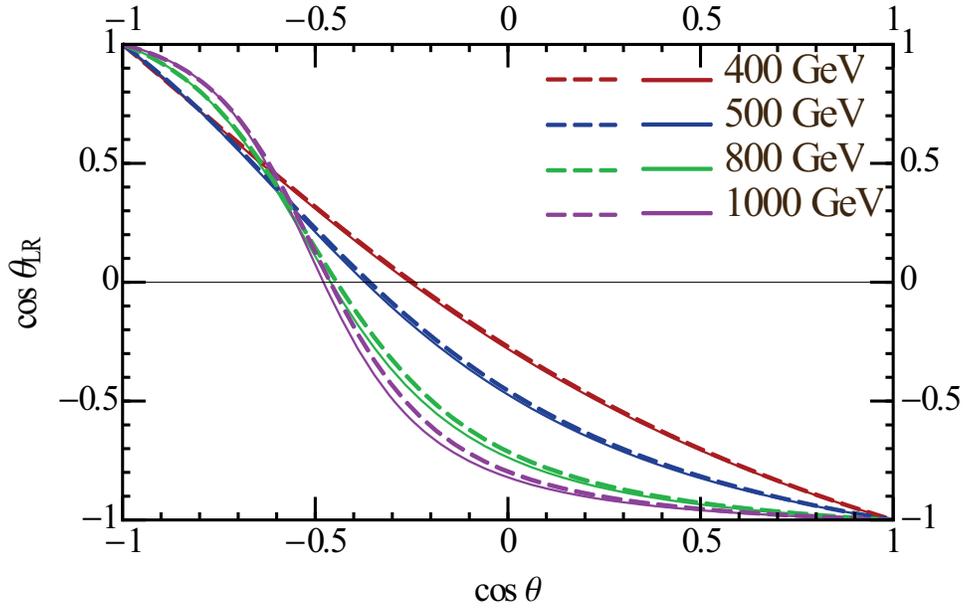, scale=0.8}
\caption{\label{costlr}$\cos\theta_{LR}$ as a function of $\cos\theta$ for
different energies $\sqrt{q^2}=400$ (red lines), $500$ (blue), $800$ (green),
and $1000\GeV$ (purple), at $\cos\theta=0$ distinguishable
from top to bottom. While the colour convention is the same as in
Figs.~\ref{figOll}--\ref{figOtra}, the line style distinguishes between Born
term results (dashed lines) and $O(\alpha_s)$ results (solid lines).}
\end{center}\end{figure}
In Fig.~\ref{costlr} we show $\cos\theta_{LR}$ as a function of $\cos\theta$
for the four center-of-mass energies used throughout this paper. The Born
term result for $\sqrt{q^2}=400\GeV$ coincides perfectly with the result shown
in Fig.~2 of Ref.~\cite{Parke:1996pr}, if we take into account that
$\xi=-\theta_{LR}$ is defined counterclockwise. Radiative corrections vanish
at the boundaries $\cos\theta=\pm 1$ and are maximal close to
$\cos\theta=-0.2$ amounting to absolute changes of $-1.0\%$
($\sqrt{q^2}=400\GeV$), $-1.9\%$ ($500\GeV$), $-3.8\%$ ($800\GeV$), and
$-4.4\%$ ($1000\GeV$).

\begin{figure}\begin{center}
\epsfig{figure=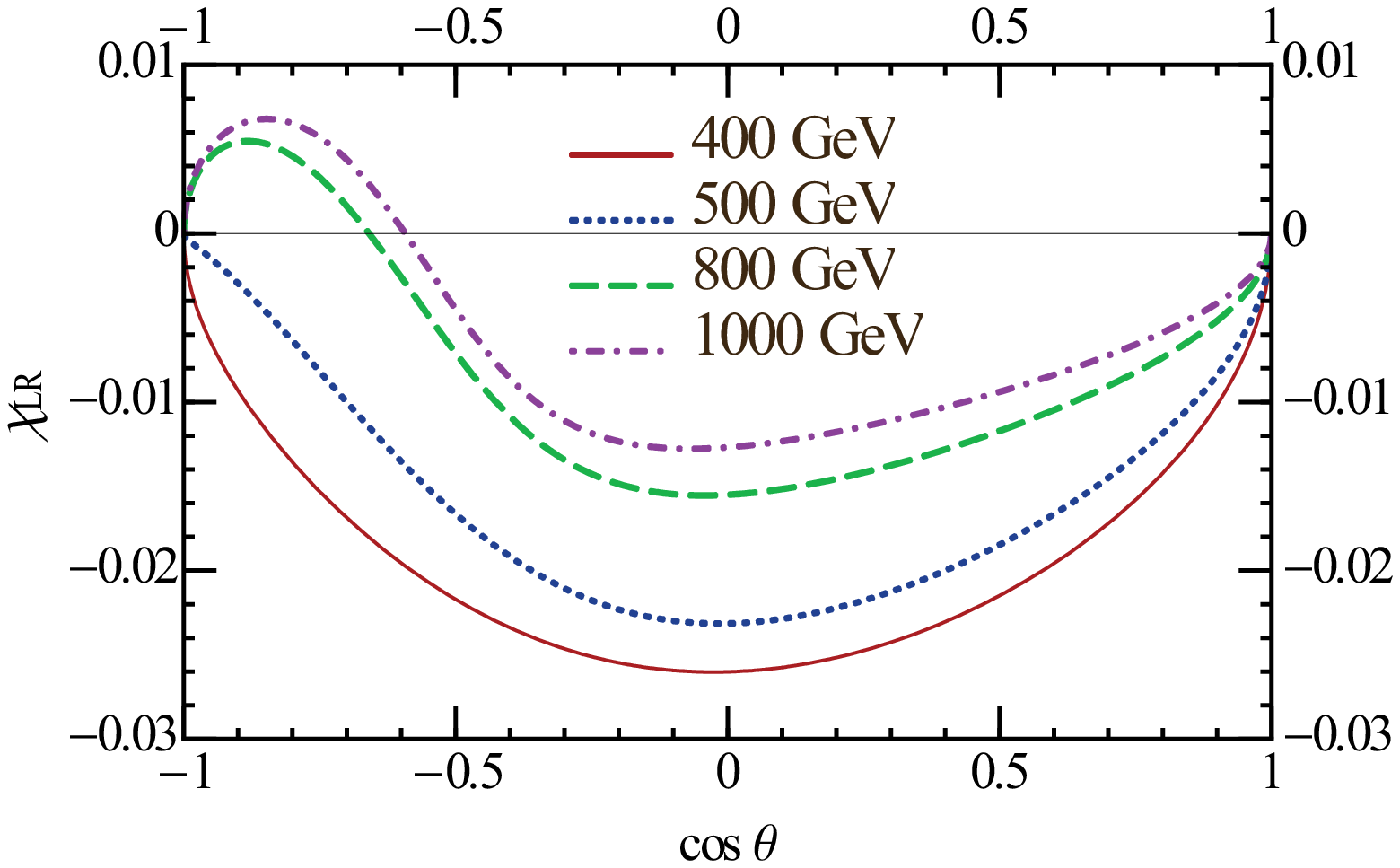, scale=0.8}
\caption{\label{xlr}$\chi_{LR}$ as a function of $\cos\theta$ for different
energies $\sqrt{q^2}=400$ (solid line), $500$ (dotted), $800$ (dashed), and
$1000\GeV$ (dashed dotted). Born term results are 2 orders of magnitude
smaller and coincide with the abscissa.}
\end{center}\end{figure}

While the irrelevance of the $Z$ width far from the $Z$ pole is nicely
demonstrated by the fact that the absolute value of the azimuthal angle
$\chi_{LR}$ at the Born term level is below $0.00035$, due to the three-body
kinematics of the final state the value for the azimuthal angle is 2 orders
of magnitude higher if one includes $O(\alpha_s)$ radiative corrections. Note
that $\chi_{LR}$ is an odd function of $\theta$. Therefore, in Fig.~\ref{xlr}
$\chi_{LR}$ is shown again as a function of $\cos\theta$ only.

\begin{figure}\begin{center}
\epsfig{figure=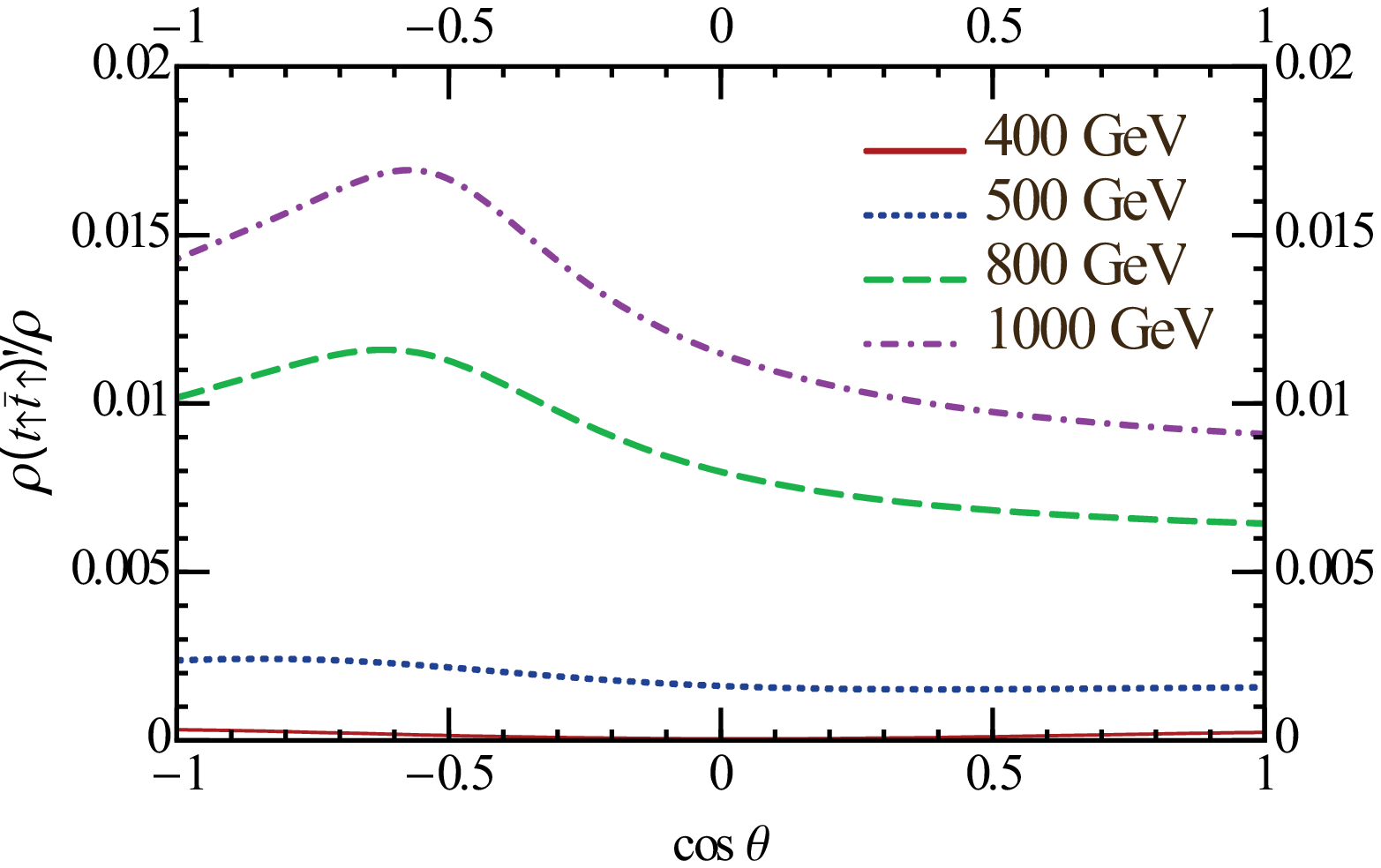, scale=0.8}
\caption{\label{offdia}$\rho(t_\uparrow\bar t_\uparrow)'/\rho$ as a function
of $\cos\theta$ for different energies $\sqrt{q^2}=400$ (solid line), $500$
(dotted), $800$ (dashed), and $1000\GeV$ (dashed dotted)}
\end{center}\end{figure}

With $\theta_{LR}$ and $\chi_{LR}$ at hand, one finally can calculate the
value for $\rho(t_\uparrow\bar t_\uparrow)'$. While this value is zero at
the Born term level, with
\begin{equation}
\vec e^{\,l\,'}=\sin\theta_{LR}(\cos\chi_{LR}\vec e^{\,t}
  +\sin\chi_{LR}\vec e^{\,n})+\cos\theta_{LR}\vec e^{\,l}
\end{equation}
and using Eqs.~(\ref{tutu}) and~(\ref{rhoee}) one obtains
\begin{eqnarray}
\rho(t_\uparrow\bar t_\uparrow)'&=&\frac14\Big(\rho
  -\rho^{tt}\sin^2\theta_{LR}\cos^2\chi_{LR}
  -\rho^{nn}\sin^2\theta_{LR}\sin^2\chi_{LR}
  -\rho^{ll}\cos^2\theta_{LR}\nonumber\\&&\strut
  -2\rho^{tn}\sin^2\theta_{LR}\sin\chi_{LR}\cos\chi_{LR}
  -2\rho^{tk}\sin\theta_{LR}\cos\theta_{LR}\cos\chi_{LR}
  \nonumber\\[3pt]&&\strut
  -2\rho^{nk}\sin\theta_{LR}\cos\theta_{LR}\sin\chi_{LR}\Big).
\end{eqnarray}
The result for the normalised quantity
$\rho(t_\uparrow\bar t_\uparrow)'/\rho$ is shown in Fig.~\ref{offdia}.
For $\sqrt{q^2}=400\GeV$ the result is still very small with a minimum value
of $0.005\%$ close to $\cos\theta=0$. The deviation grows for higher
center-of-mass energies, and a maximum is found again close to
$\cos\theta=-0.67$ with values of $\rho(t_\uparrow\bar t_\uparrow)'/\rho=1.2\%$
and $1.7\%$ for $\sqrt{q^2}=800$ and $1000\GeV$, respectively. This
observation again confirms the rigidity of the back-to-back
direction~\cite{Groote:1998xc}.

\section{Summary and conclusion}
We have presented the results of an analytical $O(\alpha_s)$ calculation of
polarised top-antitop quark production in $e^+e^-$ annihilation within the
Standard Model. We have checked our results against previously available
analytical $O(\alpha_s)$ results on single-spin polarisation effects as well
as previous numerical $O(\alpha_s)$ results on spin--spin polarisation effects
which were obtained with the phase space slicing method. Our results were
presented in the form of spin--spin density matrices defined in the respective
rest frames of the top quark and antitop quark. Based on the spin--spin
density formalism we discussed how the spin--spin correlations can be measured
through an angular analysis of the polarised top quark decays
$t(\uparrow)\to bW^+(\to\ell^++\nu)$ and the corresponding antitop quark
decay. We have briefly discussed how to generalise our results to the case of
polarised $e^+e^-$ annihilation which has allowed us to discuss the
$O(\alpha_s)$ corrections to the LO maximal spin-spin correlation effects in
the off-diagonal basis which were discovered by Parke and
Shadmi~\cite{Parke:1996pr}.

Our results have been obtained in the so-called beam frame defined by the
incoming beam electron and the outgoing top quark. Corresponding $O(\alpha_s)$
results for the so-called event frame spanned by the top--antitop quark and
the gluon will be presented in a forthcoming publication~\cite{Kaldamae:2016}.

\subsection*{Acknowledgments}
This work was supported by the Estonian Research Council under Grant
No.~IUT2-27. S.G.\ acknowledges support by the Mainz Institute of
Theoretical Physics (MITP).

\newpage

\begin{appendix}

\section{Dilogarithmic integrals}
\setcounter{equation}{0}\def\theequation{A\arabic{equation}}
This appendix contains the standard logarithms and the dilogarithmic integrals
which are the main building blocks for the $O(\alpha_s)$ final results. The
logarithms necessary to write up the results are given by
\begin{eqnarray}
&\displaystyle
\ell_0^-=\ln\pfrac{4(1-\sxi)}\xi\qquad
\ell_0^+=\ln\pfrac{2(1+\sxi)}\sxi\nonumber\\
&\displaystyle
\ell_1=\ln\pfrac{\sxi(1+\sxi)}2\qquad
\ell_2=\ln\pfrac{2-\sxi}\sxi\qquad
\ell_3=\ln\pfrac{1+v}{1-v}.&
\end{eqnarray}
The dilogarithmic integrals are given by Eqs.~(\ref{Cintdef}) for $n=0$. A
special treatment is necessary for the integral $I^{L_0}_{t-}(0)$ because this
integral is IR divergent. After extracting the divergent part according to the
method explained in the main text, one is left with the subtracted integral
\begin{eqnarray}
I^{\prime L_0}_{t-}(0)&=&\int_{t_-}^{t_1}\frac{L_0(t)dt}{t-t_-}
  -2\ln\pfrac{1+v}{1-v}\int_{t_-}^{t_1}
  \left(\frac{dt}{t-t_-}+\frac{dt}{t-t_+}-\frac{dt}t\right)\ =\nonumber\\
  &=&-\ln\pfrac{1-v}{2(1+\sxi)}\ln\pfrac{1+v}{1-v}
  -\Li_2\left(-\frac{1-v-\sxi}{1+v}\right)\strut\nonumber\\&&\strut
  +\Li_2\pfrac{1-v-\sxi}{1-v}+\Li_2\left(2\frac{1-v-\sxi}{(1-v)^2}\right).
\end{eqnarray}
As the IR divergence is general, the coefficient for this integral is also
proportional to the Born term result. As such, one can resum the contributions
from the IR-divergent part of the tree-term diagrams,
\begin{eqnarray}
I_D&=&2\ln\pfrac{1-\sxi}\sxi\ln\pfrac{1+v}{1-v}
  -\Li_2\pfrac{2v}{(1+v)^2}+\Li_2\pfrac{-2v}{(1-v)^2}\strut\nonumber\\&&\strut
  +\frac12\Li_2\left(-\frac{(1-v)^2}{(1+v)^2}\right)
  -\frac12\Li_2\left(-\frac{(1+v)^2}{(1-v)^2}\right)
\end{eqnarray}
and the contributions from the IR-divergent part of the one-loop diagrams,
\begin{equation}
I_L=\ln\pfrac\xi4\ln\pfrac{1+v}{1-v}+\Li_2\pfrac{2v}{1+v}
  -\Li_2\pfrac{-2v}{1-v}-\pi^2
\end{equation}
to obtain
\begin{equation}
\hat I^{L_0}_{t-}(0)=I^{\prime L_0}_{t-}(0)+I_D-I_L.
\end{equation}
For the other dilogarithmic integrals there is no need for regularisation.
Using the known identities for dilogarithms, one can of course try hard to
simplify the expressions in terms of dilogarithms and double logarithms.
However, the outcome is still arbitrary and in general will not justify the
effort. Still, we brought the results into a form which is manifestly real,
obtaining
\begin{eqnarray}
I^{L_0}_{t+}(0)&=&\ln\pfrac{2(1+\sxi)}{1+v}\ln\pfrac{1+v}{1-v}
  +\Li_2\pfrac{-2v}{1-v}-\Li_2\pfrac{2v}{1+v}-\Li_2\pfrac{4v}{(1+v)^2}
  \strut\nonumber\\&&\strut
  -\Li_2\left(-\frac{1+v-\sxi}{1-v}\right)
  +\Li_2\pfrac{1+v-\sxi}{1+v}+\Li_2\left(2\frac{1+v-\sxi}{(1+v)^2}\right),
  \nonumber\\[12pt]
I^{L_0}_{1-}(0)&=&-\Li_2\pfrac{1-v-\sxi}{2-\sxi}
  +\Li_2\left(-\frac{1-v-\sxi}\sxi\right)
  +\Li_2\left(-2\frac{1-v-\sxi}{\sxi(2-\sxi)}\right),\nonumber\\[12pt]
I^{L_0}_{1+}(0)&=&\frac12\ln\pfrac{2(1+v)}{1+\sxi}\ln\pfrac{2+\sxi}2
  +\frac14\ln\pfrac{2(1+\sxi)}{1+v}^2\strut\nonumber\\&&\strut
  -\ln^22-\Li_2\pfrac{2\sxi}{2+\sxi}-\Li_2\pfrac12-\Li_2\pfrac{2+\sxi}4
  \strut\nonumber\\&&\strut
  +\Li_2\pfrac{1-v+\sxi}{2+\sxi}+\Li_2\pfrac{\sxi}{1-v+\sxi}
  +\Li_2\pfrac{\sxi(2+\sxi)}{2(1-v+\sxi)},\nonumber\\[12pt]
I^{L_0}_0(0)&=&\frac14\ln\pfrac{1+v}{1-v}^2+\Li_2\pfrac{1-v}2
  +\Li_2\pfrac{1+v}2-2\Li_2\pfrac\sxi2,\nonumber\\[12pt]
I^{L_+}_{t-}(0)&=&2\Li_2\left(-\frac{1-v-\sxi}{1+v+\sxi}\right)
  -2\Li_2\pfrac{(2+\sxi)(1-v-\sxi)}{(1-v)(1-v+\sxi)}\strut\nonumber\\&&\strut
  -\Li_2\left(-\frac{1-v-\sxi}{1+v}\right)+\Li_2\pfrac{1-v-\sxi}{1-v}
  +\Li_2\pfrac{2(1-v-\sxi)}{(1-v)^2},\nonumber\\[12pt]
I^{L_+}_{t+}(0)&=&\Li_2\pfrac{-2v}{1-v}-\Li_2\pfrac{2v}{1+v}
  -\Li_2\pfrac{4v}{(1+v)^2}\strut\nonumber\\&&\strut
  +\Li_2\pfrac{2(1+v-\sxi)}{(1+v)^2}-\Li_2\left(-\frac{1+v-\sxi}{1-v}\right)
  +\Li_2\pfrac{1+v-\sxi}{1+v}\strut\nonumber\\&&\strut
  -2\Li_2\pfrac{(2+\sxi)(1+v-\sxi)}{(1+v)(1+v+\sxi)}
  +2\Li_2\pfrac{2(2+\sxi)v}{(1+v)(1+v+\sxi)}\strut\nonumber\\&&\strut
  +2\Li_2\left(-\frac{1+v-\sxi}{1-v+\sxi}\right)
  -2\Li_2\pfrac{-2v}{1-v+\sxi},\nonumber\\[12pt]
I^{L_+}_{1-}(0)&=&-\Li_2\pfrac{1-v-\sxi}{2-\sxi}
  +\Li_2\left(-\frac{1-v-\sxi}\sxi\right)
  +\Li_2\left(-2\frac{1-v-\sxi}{\sxi(2-\sxi)}\right)\strut\nonumber\\&&\strut
  -2\Li_2\left(-\frac{(2+\sxi)(1-v-\sxi)}{2\sxi}\right)
  +2\Li_2\pfrac{1-v-\sxi}2,\nonumber\\[12pt]
I^{L_+}_{1+}(0)&=&\frac12\ln\pfrac{(2+\sxi)^3}{8(1+\sxi)^2}
  \ln\pfrac{2(1+v)}{1+\sxi}-\Li_2\pfrac{2\sxi}{2+\sxi}-\Li_2\pfrac12
  \strut\nonumber\\&&\strut
  -\Li_2\pfrac{2+\sxi}4+\Li_2\pfrac{1-v+\sxi}{2+\sxi}
  +\Li_2\pfrac\sxi{1-v+\sxi}\strut\nonumber\\&&\strut
  +\Li_2\pfrac{\sxi(2+\sxi)}{2(1-v+\sxi)}
  +2\Li_2\pfrac{1+\sxi}{2+\sxi}+2\Li_2\pfrac\sxi{1+\sxi}
  \strut\nonumber\\&&\strut
  -2\Li_2\pfrac{2\sxi(1+\sxi)}{(2+\sxi)(1-v+\sxi)}
  -2\Li_2\pfrac{1-v+\sxi}{2(1+\sxi)},\nonumber\\[12pt]
I^{L_+}_0(0)&=&4\Li_2\pfrac\sxi{2+\sxi}-2\Li_2\pfrac{1+v}{2+\sxi}
  -2\Li_2\pfrac{1-v}{2+\sxi}\strut\nonumber\\&&\strut
  +\Li_2\pfrac{1-v}2+\Li_2\pfrac{1+v}2-2\Li_2\pfrac\sxi2.
\end{eqnarray}
Still, not all of these integrals are independent of each other. Because of
\begin{equation}
\frac{dy}y=\left(\frac1{t-t_-}+\frac1{t-t_+}-\frac1t\right)dt
\end{equation}
for the standard substitution $t=(1-y-\sqrt{(1-y)^2-\xi})/\sxi$ used in this
paper, and
\begin{equation}
L_0=\int_{z_-}^{z_+}\frac{dz}z,\qquad
L_\pm=-\int_{z_-}^{z_+}\frac{dz}{1\pm\sxi-z},
\end{equation}
using the symmetry of the phase space for $t\bar t$ pairs, one obtains
\begin{eqnarray}\label{rel1}
\lefteqn{I^{L_\pm}_{t-}(0)+I^{L_\pm}_{t+}(0)-I^{L_\pm}_0(0)
  \ =\ \int_0^{1-\sxi}\int_{z_-(y)}^{z_+(y)}\frac{dz}{1\pm\sxi-z}\,\frac{dy}y}
  \nonumber\\
  &=&\int_0^{1-\sxi}\int_{z_-(y)}^{z_+(y)}\frac{dz}z\,\frac{dy}{1\pm\sxi-y}
  \ =\ I^{L_0}_0(0)\mp 2I^{L_0}_{1\pm}(0),
\end{eqnarray}
where for the last step we used
\begin{equation}
\frac{dy}{1\pm\sxi-y}=\left(\frac1t\mp\frac2{1\pm t}\right)dt.
\end{equation}
Using Eq.~(\ref{rel1}) to eliminate $I^{L_0}_{1+}(0)$, we obtain the results
presented in Appendix~B. The relation also takes into account the integrals
with $L_-(t)$ which are absent in our case. If such integrals appear in an
intermediate step, they can be eliminated by using the relation
\begin{eqnarray}
\lefteqn{I^{L_+}_0(0)+2I^{L_+}_{1-}(0)
  \ =\ -\int_0^{1-\sxi}\int_{z_-(y)}^{z_+(y)}
  \frac{dz}{1+\sxi-z}\,\frac{dy}{1-\sxi-y}}\nonumber\\
  &=&-\int_0^{1-\sxi}\int_{z_-(y)}^{z_+(y)}\frac{dz}{1-\sxi-z}\,
  \frac{dy}{1+\sxi-y}\ =\ I^{L_-}_0(0)-2I^{L_-}_{1+}(0).
\end{eqnarray}
The nonappearance of $L_-(t)$ is related to the fact that there is no
singularity at the upper boundary $y=1-\sxi$.

\section{$O(\alpha_s)$ results}
\setcounter{equation}{0}\def\theequation{B\arabic{equation}}
In this appendix we present our analytic results for the $O(\alpha_s)$
contributions to the correlation matrix in terms of the three unit vectors
$\hat t$, $\hat n$ and $\hat l$ in the laboratory frame. The detailed results
have to be combined with the electroweak form factors (cf.\
Ref.~\cite{Groote:1996nc}) to obtain
\begin{equation}
\rho^{P_1P_2}=\sum_{i,j=1}^4g_{ij}\rho^{P_1P_2}_{ij}.
\end{equation}
($P_1,P_2\in\{t,n,l\}$). The general factor $N$ is given in Eq.~(\ref{genfac}).
Again, $\rho^{P_1P_2}_{ij}$ is divided up into five different angular
dependences,
\begin{eqnarray}
\rho^{P_1P_2}_{ij}&=&\frac14(1+\cos^2\theta)\rho^{P_1P_2}_{ijU}
  +\frac12\sin^2\theta\rho^{P_1P_2}_{ijL}
  +\frac12\cos\theta\rho^{P_1P_2}_{ijF}\nl
  +\frac12\sin\theta\cos\theta\rho^{P_1P_2}_{ijI}
  +\frac12\sin\theta\rho^{P_1P_2}_{ijA},
\end{eqnarray}
where the additional indices stand for unpolarised transverse ($U$),
longitudinal ($L$), forward-/backward -symmetric ($F$),
longitudinal/transverse interference ($I$), and parity-asymmetric ($A$)
components of the intermediate ($\gamma$ or $Z$) boson. Our results read
\input tnlalpha

\end{appendix}


\end{document}

%% file: tnlalpha.tex
\begin{eqnarray}
\lefteqn{\rho^{tt}_{11U}\ =\ N\bigg[
  \frac{2v}{(1+\sxi)\sxi}(4+8\sxi+13\xi+13\xi^{3/2})}\nl
  +\frac1{(1+\sxi)\sxi}(4+4\sxi-\xi-23\xi^{3/2}-12\xi^2-2\xi^{5/2})\ell_3\nl
  +\frac\sxi{4(1+\sxi)^2}(176+369\sxi+320\xi+134\xi^{3/2}+24\xi^2+\xi^{5/2})
  I^{L_+}_0(0)\nl
  +\frac8{(1+\sxi)v^2\xi}(1+3\sxi-\xi-\xi^{3/2})I^{L_+}_{1-}(0)
  +\frac8\xi(1+\sxi)^4(1-2\sxi)I^{L_+}_{1+}(0)\nl
  +\frac{(1+\sxi)\sxi}{4(1-\sxi)}(16-\sxi-12\xi+\xi^{3/2})\tintpp\nl
  -\frac{(1-\sxi)\sxi}{4(1+\sxi)^2}(16+\sxi+\xi+7\xi^{3/2}-\xi^2)
  \left(I^{L_0}_0(0)+2I^{L_0}_{1-}(0)\right)\bigg],\\[12pt]
\lefteqn{\rho^{tt}_{11L}\ =\ N\bigg[
  -\frac1{2v\sxi}(20+20\sxi+65\xi+2\xi^{3/2}-105\xi^2-18\xi^{5/2})}\nl
  -\frac1{4v^2\sxi}(20-101\xi-166\xi^{3/2}+143\xi^2+196\xi^{5/2}-2\xi^3
  -42\xi^{7/2})\ell_3\nl
  -\frac\sxi{16(1+\sxi)^2}(896+2033\sxi+1932\xi+1006\xi^{3/2}+300\xi^2
  +41\xi^{5/2})I^{L_+}_0(0)\nl
  -\frac2{v^4\xi}(5+10\sxi-19\xi+12\xi^2)I^{L_+}_{1-}(0)
  -\frac2\xi(1+\sxi)^4(5-10\sxi+\xi)I^{L_+}_{1+}(0)\nl
  -\frac\xi{16(1-\sxi)^2}(47-72\sxi+122\xi-8\xi^{3/2}-57\xi^2)\tintpp\nl
  -\frac\xi{16(1+\sxi)^2}(271+196\sxi-78\xi-92\xi^{3/2}-41\xi^2)I^{L_0}_0(0)\nl
  -\frac{(1-\sxi)\xi}{8(1+\sxi)^2}(143+83\sxi-59\xi-23\xi^{3/2})
  I^{L_0}_{1-}(0)\nl
  +4\xi\tintsp\bigg],\\[12pt]
\lefteqn{\rho^{tt}_{12U}\ =\ N\bigg[
  4v(6+3\sxi+4\xi)-2\sxi(2+4\sxi+\xi)\ell_3}\nl
  +\frac{2+\sxi}{1+\sxi}(8+12\sxi+18\xi+11\xi^{3/2}+3\xi^2)I^{L_+}_0(0)\nl
  +\frac{24}{v^2}I^{L_+}_{1-}(0)-8(1+\sxi)^4I^{L_+}_{1+}(0)
  +\frac{(1+\sxi)\xi}{1-\sxi}(4-\xi)\tintpp\nl
  +\frac{\xi^{3/2}}{1+\sxi}(1-\sxi)(2-\sxi)
  \left(I^{L_0}_0(0)+2I^{L_0}_{1-}(0)\right)\bigg],\\[12pt]
\lefteqn{\rho^{tt}_{12U}\ =\ N\bigg[
  \frac{2\sxi}{v}(12+3\sxi-12\xi+\xi^{3/2})}\nl
  +\frac1{v^2}(20-4\sxi-46\xi+2\xi^{3/2}+11\xi^2+2\xi^{5/2}+3\xi^3)\ell_3\nl
  -\frac1{2(1+\sxi)^2}(16+16\sxi-20\xi-59\xi^{3/2}-49\xi^2-15\xi^{5/2}-\xi^3)
  I^{L_+}_0(0)\nl
  -\frac4{v^4}(5-\xi)I^{L_+}_{1-}(0)-4(1+\sxi)^4I^{L_+}_{1+}(0)\nl
  -\frac\xi{2(1-\sxi)^2}(16-13\sxi+\xi+3\xi^{3/2}-3\xi^2)\tintpp\nl
  -\frac1{2(1+\sxi)^2}(16+32\sxi+4\xi-9\xi^{3/2}-7\xi^2-5\xi^{5/2}+\xi^3)
  I^{L_0}_0(0)\nl
  +\frac{(1-\sxi)\xi}{(1+\sxi)^2}(4-3\sxi-4\xi+\xi^{3/2})I^{L_0}_{1-}(0)\nl
  +4\tintsp\bigg],\\[12pt]
\lefteqn{\rho^{tt}_{44F}\ =\ N\bigg[
  -\frac{4\sxi}{1+\sxi}(6-8\sxi-5\xi+3\xi^{3/2})
  -\frac4{v^2}(4-4\sxi+3\xi^{3/2}+\xi^2)\ell_1}\nl
  -\frac{2(1+\sxi)}{1-\sxi}(4-\xi)(2-\sxi)\ell_2
  +\frac{2v\sxi}{(1+\sxi)^2}(20+30\sxi+19\xi+9\xi^{3/2}+3\xi^2)\ell_3\nl
  +\frac{(2+\sxi)^2\sxi}{(1+\sxi)^2}(4+3\sxi+2\xi)I^{L_+}_0(0)
  -\frac{v\xi^2}{(1-\sxi)^2}\tintpp\nl
  -\frac\sxi{(1+\sxi)^2}(8+12\sxi+24\xi+27\xi^{3/2}+10\xi^2)
  I^{L_0}_0(0)\bigg],\\[12pt]
\lefteqn{\rho^{tn}_{13U}\ =\ N\bigg[
  -\frac4{1+\sxi}(6-3\sxi-5\xi+4\xi^{3/2}-6\xi^2)}\nl
  -\frac4{v^2}(12+8\sxi-4\xi-9\xi^{3/2}-3\xi^2)\ell_1\nl
  -\frac2{1-\sxi}(24-16\sxi+6\xi-10\xi^{3/2}-\xi^2+3\xi^{5/2})\ell_2\nl
  -\frac{2v\xi}{(1+\sxi)^2}(6-\sxi-\xi+3\xi^{3/2})\ell_3\nl
  -\frac{2+\sxi}{(1+\sxi)^2}(8+20\sxi+30\xi+21\xi^{3/2}+6\xi^2)I^{L_+}_0(0)\nl
  -\frac{v\xi}{(1-\sxi)^2}(4-3\xi)\tintpm
  -\frac{\xi^{3/2}}{(1+\sxi)^2}(6-5\sxi-8\xi)I^{L_0}_0(0)\bigg],\\[12pt]
\lefteqn{\rho^{tn}_{13L}\ =\ N\bigg[
  -\frac{2\sxi}{1+\sxi}(16-2\sxi-21\xi+3\xi^{3/2})
  -\frac2{v^2}(4+12\sxi-4\xi-7\xi^{3/2}-\xi^2)\ell_1}\nl
  +\frac1{1-\sxi}(8-28\sxi+10\xi+5\xi^{3/2}-\xi^2)\ell_2\nl
  +\frac{v}{(1+\sxi)^2}(4+8\sxi-2\xi-23\xi^{3/2}-11\xi^2+3\xi^{5/2})\ell_3\nl
  -\frac{(2+\sxi)^2}{2(1+\sxi)^2}(4+16\sxi+17\xi+8\xi^{3/2})I^{L_+}_0(0)\nl
  -\frac{v\xi}{2(1-\sxi)^2}(12-4\sxi-7\xi)\tintpm\nl
  -\frac1{2(1+\sxi)^2}(16+32\sxi-8\xi-52\xi^{3/2}-17\xi^2+8\xi^{5/2})
  I^{L_0}_0(0)\nl
  +4v\tintsm\bigg],\\[12pt]
\lefteqn{\rho^{tn}_{14L}\ =\ 4N\pi v\xi,}\\[12pt]
\lefteqn{\rho^{tl}_{11I}\ =\ N\bigg[
  -\frac1{v\sxi}(16+16\sxi+20\xi-12\xi^{3/2}-55\xi^2-4\xi^{5/2}+3\xi^3)}\nl
  -\frac1{2v^2\sxi}(16-4\xi-24\xi^{3/2}+26\xi^2+28\xi^{5/2}+7\xi^3-4\xi^{7/2}
  +3\xi^4)\ell_3\nl
  -\frac\sxi{2(1+\sxi)^2}(152+303\sxi+211\xi+51\xi^{3/2}-3\xi^2-2\xi^{5/2})
  I^{L_+}_0(0)\nl
  -\frac8{v^4\xi}(2+4\sxi-7\xi+5\xi^2)I^{L_+}_{1-}(0)
  -\frac8\xi(1+\sxi)^4(2-4\sxi+\xi)I^{L_+}_{1+}(0)\nl
  -\frac\sxi{2(1-\sxi)^2}(8+25\sxi-39\xi+9\xi^{3/2}+11\xi^2-6\xi^{5/2})
  \left(I^{L_+}_{t-}(0)+I^{L_+}_{t+}(0)\right)\nl
  +\frac\sxi{2(1+\sxi)^2}(24+15\sxi+9\xi-\xi^{3/2}-13\xi^2-2\xi^{5/2})
  I^{L_0}_0(0)\nl
  +\frac{(1-\sxi)^2\sxi}{(1+\sxi)^2}(8-\sxi-9\xi-2\xi^{3/2})I^{L_0}_{1-}(0)\nl
  -4\sxi\left((2-\xi)\left(\hat I^{L_0}_{t-}(0)+I^{L_0}_{t+}(0)\right)
  -4v(\ell_0^-+\ell_0^+)\right)\bigg],\\[12pt]
\lefteqn{\rho^{tl}_{12I}\ =\ N\bigg[
  -\frac1{v}(64-4\sxi-84\xi+7\xi^{3/2}+4\xi^2-3\xi^{5/2})}\nl
  -\frac\sxi{2v^2}(40+64\sxi-74\xi-20\xi^{3/2}+37\xi^2+4\xi^{5/2}-3\xi^3)
  \ell_3\nl
  -\frac1{2(1+\sxi)^2}(64+160\sxi+212\xi+159\xi^{3/2}+65\xi^2+17\xi^{5/2}
  +3\xi^3)I^{L_+}_0(0)\nl
  -\frac8{v^4}(7-3\xi)I^{L_+}_{1-}(0)+8(1+\sxi)^4I^{L_+}_{1+}(0)\nl
  -\frac\xi{2(1-\sxi)^2}(52-31\sxi-29\xi+11\xi^{3/2}+5\xi^2)
  \left(I^{L_+}_{t-}(0)+I^{L_+}_{t+}(0)\right)\nl
  +\frac\sxi{2(1+\sxi)^2}(16+52\sxi-3\xi-35\xi^{3/2}-\xi^2+3\xi^{5/2})
  I^{L_0}_0(0)\nl
  +\frac{(1-\sxi)\xi}{(1+\sxi)^2}(20+9\sxi-10\xi-3\xi^{3/2})I^{L_0}_{1-}(0)\nl
  -4\sxi\left((2-\xi)\left(\hat I^{L_0}_{t-}(0)+I^{L_0}_{t+}(0)\right)
  -4v(\ell_0^-+\ell_0^+)\right)\bigg],\\[12pt]
\lefteqn{\rho^{tl}_{43A}\ =\ 2N\pi v\sxi(1+\xi),}\\[12pt]
\lefteqn{\rho^{tl}_{44A}\ =\ N\bigg[
  \frac\sxi{1+\sxi}(24-10\sxi-7\xi+6\xi^{3/2}+3\xi^2)
  +\frac4{v^2}(4-5\xi^{3/2}-3\xi^2)\ell_1}\nl
  +\frac1{2(1-\sxi)}(32+8\sxi-80\xi-6\xi^{3/2}+22\xi^2+3\xi^{5/2}-3\xi^3)
  \ell_2\nl
  -\frac{2v\sxi}{(1+\sxi)^2}(21+32\sxi+12\xi-5\xi^{3/2}-3\xi^2)\ell_3\nl
  -\frac{(2+\sxi)\sxi}{(1+\sxi)^2}(8+10\sxi+5\xi)I^{L_+}_0(0)\nl
  -\frac{v\xi}{(1-\sxi)^2}(8-4\sxi-3\xi)
  \left(I^{L_+}_{t-}(0)-I^{L_+}_{t+}(0)\right)\nl
  +\frac\sxi{(1+\sxi)^2}(16+28\sxi+17\xi-\xi^{3/2}-3\xi^2)I^{L_0}_0(0)\nl
  -4v\sxi\left((2-\xi)\left(\hat I^{L_0}_{t-}(0)-I^{L_0}_{t+}(0)\right)
  -4v\ell_0^-\right)\bigg],\\[12pt]
\lefteqn{\rho^{nt}_{13U}\ =\ N\bigg[
  \frac4{1+\sxi}(6-3\sxi-5\xi+4\xi^{3/2}-6\xi^2)
  +\frac4{v^2}(12+8\sxi-4\xi-9\xi^{3/2}-3\xi^2)\ell_1}\nl
  +\frac2{1-\sxi}(24-16\sxi+6\xi-10\xi^{3/2}-\xi^2+3\xi^{5/2})\ell_2\nl
  +\frac{2v\xi}{(1+\sxi)^2}(6-\sxi-\xi+3\xi^{3/2})\ell_3\nl
  +\frac{2+\sxi}{(1+\sxi)^2}(8+20\sxi+30\xi+21\xi^{3/2}+6\xi^2)I^{L_+}_0(0)\nl
  +\frac{v\xi}{(1-\sxi)^2}(4-3\xi)\tintpm
  +\frac{\xi^{3/2}}{(1+\sxi)^2}(6-5\sxi-8\xi)I^{L_0}_0(0)\bigg],\\[12pt]
\lefteqn{\rho^{nt}_{13L}\ =\ N\bigg[
  \frac{2\sxi}{1+\sxi}(20+10\sxi-19\xi+9\xi^{3/2})
  -\frac2{v^2}(20-4\sxi-4\xi+5\xi^{3/2}+3\xi^2)\ell_1}\nl
  -\frac1{(1-\sxi)}(24-36\sxi+46\xi+3\xi^{3/2}-7\xi^2)\ell_2\nl
  +\frac{v}{(1+\sxi)^2}(4+8\sxi-42\xi-51\xi^{3/2}-23\xi^2-9\xi^{5/2})\ell_3\nl
  -\frac{(2+\sxi)^2}{2(1+\sxi)^2}(4+5\xi)I^{L_+}_0(0)
  -\frac{v\xi}{2(1-\sxi)^2}(12-4\sxi-3\xi)\tintpm\nl
  -\frac1{2(1+\sxi)^2}(16+32\sxi-24\xi-60\xi^{3/2}-53\xi^2-24\xi^{5/2})
  I^{L_0}_0(0)\nl
  +4v\tintsm\bigg],\\[12pt]
\lefteqn{\rho^{nt}_{14L}\ =\ 4N\pi v\xi,}\\[12pt]
\lefteqn{\rho^{nn}_{11U}\ =\ N\bigg[
  \frac{2v}{(1+\sxi)\sxi}(4+8\sxi+13\xi+13\xi^{3/2})}\nl
  +\frac1{(1+\sxi)\sxi}(4+4\sxi-\xi-23\xi^{3/2}-12\xi^2-2\xi^{5/2})\ell_3\nl
  +\frac\sxi{4(1+\sxi)^2}(176+369\sxi+320\xi+134\xi^{3/2}+24\xi^2+\xi^{5/2})
  I^{L_+}_0(0)\nl
  +\frac8{(1+\sxi)v^2\xi}(1+3\sxi-\xi-\xi^{3/2})I^{L_+}_{1-}(0)\nl
  +\frac8\xi(1+\sxi)^4(1-2\sxi)I^{L_+}_{1+}(0)\nl
  +\frac{(1+\sxi)\sxi}{4(1-\sxi)}(16-\sxi-12\xi+\xi^{3/2})\tintpp\nl
  -\frac{(1-\sxi)\sxi}{4(1+\sxi)^2}(16+\sxi+\xi+7\xi^{3/2}-\xi^2)
  \left(I^{L_0}_0(0)+2I^{L_0}_{1-}(0)\right)\bigg],\\[12pt]
\lefteqn{\rho^{nn}_{11L}\ =\ N\bigg[
  -\frac{v}{2(1+\sxi)\sxi}(12+24\sxi+79\xi+77\xi^{3/2}-2\xi^2)}\nl
  -\frac1{4(1+\sxi)\sxi}(12+12\sxi-55\xi-129\xi^{3/2}-32\xi^2+12\xi^{5/2}
  +6\xi^3)\ell_3\nl
  -\frac\sxi{16(1+\sxi)^2}(640+1463\sxi+1396\xi+650\xi^{3/2}+132\xi^2
  +7\xi^{5/2})I^{L_+}_0(0)\nl
  -\frac2{(1+\sxi)v^2\xi}(3+9\sxi-4\xi-4\xi^{3/2})I^{L_+}_{1-}(0)
  -\frac2\xi(1+\sxi)^4(3-6\sxi-\xi)I^{L_+}_{1+}(0)\nl
  -\frac{(1+\sxi)\xi}{16(1-\sxi)}(41-24\sxi-9\xi)\tintpp\nl
  -\frac{(1-\sxi)\xi}{16(1+\sxi)^2}(73+5\sxi-37\xi+7\xi^{3/2})
  \left(I^{L_0}_0(0)+2I^{L_0}_{1-}(0)\right)\bigg],\\[12pt]
\lefteqn{\rho^{nn}_{12U}\ =\ N\bigg[
  4v(6+3\sxi+4\xi)-\sxi(2+4\sxi+\xi)\ell_3}\nl
  +\frac{2+\sxi}{1+\sxi}(8+12\sxi+18\xi+11\xi^{3/2}+3\xi^2)I^{L_+}_0(0)\nl
  +\frac{24}{v^2}I^{L_+}_{1-}(0)-8(1+\sxi)^4I^{L_+}_{1+}(0)
  -\frac{(1+\sxi)\xi}{1-\sxi}(4-\xi)\tintpp\nl
  +\frac{\xi^{3/2}}{1+\sxi}(1-\sxi)(2-\sxi)
  \left(I^{L_0}_0(0)+2I^{L_0}_{1-}(0)\right)\bigg],\\[12pt]
\lefteqn{\rho^{nn}_{12L}\ =\ N\bigg[
  2v\sxi(8-7\sxi)-(20+4\sxi-42\xi+4\xi^{3/2}+13\xi^2)\ell_3}\nl
  +\frac1{2(1+\sxi)}(16+64\sxi+100\xi+61\xi^{3/2}+9\xi^2-2\xi^{5/2})
  I^{L_+}_0(0)\nl
  +\frac4{v^2}I^{L_+}_{1-}(0)-12(1+\sxi)^4I^{L_+}_{1+}(0)
  +\frac{(1+\sxi)\xi}{2(1-\sxi)}(24-9\sxi-14\xi)\tintpp\nl
  +\frac{1-\sxi}{2(1+\sxi)}(16+32\sxi-4\xi-21\xi^{3/2}-2\xi^2)I^{L_0}_0(0)\nl
  -\frac{(1-\sxi)\xi}{1+\sxi}(12+5\sxi-6\xi)I^{L_0}_{1-}(0)\nl
  -4v^2\tintsp\bigg],\\[12pt]
\lefteqn{\rho^{nn}_{44F}\ =\ N\bigg[
  -\frac{4\sxi}{1+\sxi}(6-8\sxi-5\xi+3\xi^{3/2})
  -\frac4{v^2}(4-4\sxi+3\xi^{3/2}+\xi^2)\ell_1}\nl
  -\frac{2(1+\sxi)}{1-\sxi}(8-4\sxi-2\xi+\xi^{3/2})\ell_2
  +\frac{2v\sxi}{(1+\sxi)^2}(20+30\sxi+19\xi+9\xi^{3/2}+3\xi^2)\ell_3\nl
  +\frac\sxi{(1+\sxi)^2}(16+28\sxi+24\xi+11\xi^{3/2}+2\xi^2)I^{L_+}_0(0)\nl
  -\frac\sxi{(1+\sxi)^2}(8+12\sxi+24\xi+27\xi^{3/2}+10\xi^2)I^{L_0}_0(0)\nl
  -\frac{v\xi^2}{(1-\sxi)^2}\left(I^{L_+}_{t-}(0)-I^{L_+}_{t+}(0)\right)
  \bigg],\\[12pt]
\lefteqn{\rho^{nl}_{13I}\ =\ N\bigg[
  -\frac1{1+\sxi}(64-52\sxi-114\xi+45\xi^{3/2}-26\xi^2+3\xi^{5/2})}\nl
  -\frac4{v^2}(28+8\sxi-8\xi-7\xi^{3/2}-\xi^2)\ell_1\nl
  -\frac1{2(1-\sxi)}(224-168\sxi+48\xi+14\xi^{3/2}+2\xi^2+3\xi^{5/2}-3\xi^3)
  \ell_2\nl
  -\frac{2v\sxi}{(1+\sxi)^2}(1+40\sxi+32\xi+7\xi^{3/2}+5\xi^2)\ell_3\nl
  -\frac{2+\sxi}{(1+\sxi)^2}(16+32\sxi+50\xi+33\xi^{3/2}+8\xi^2)I^{L_+}_0(0)\nl
  +\frac\sxi{(1+\sxi)^2}(8+36\sxi+17\xi+11\xi^{3/2}+13\xi^2)I^{L_0}_0(0)\nl
  -\frac{v\xi}{(1-\sxi)^2}(16-4\sxi-7\xi)
  \left(I^{L_+}_{t-}(0)-I^{L_+}_{t+}(0)\right)\nl
  -4v\sxi\left((2-\xi)\left(\hat I^{L_0}_{t-}(0)-I^{L_0}_{t+}(0)\right)
  -4v\ell_0^-\right)\bigg],\\[12pt]
\lefteqn{\rho^{nl}_{14I}\ =\ -2N\pi v\sxi(1+\xi),}\\[12pt]
\lefteqn{\rho^{nl}_{41A}\ =\ -2N\pi v^2\sxi,}\\[12pt]
\lefteqn{\rho^{nl}_{42A}\ =\ -2N\pi v^2\sxi,}\\[12pt]
\lefteqn{\rho^{lt}_{11I}\ =\ N\bigg[
  \frac1{v}(48+16\sxi-40\xi+3\xi^{3/2}-32\xi^2-3\xi^{5/2})}\nl
  -\frac\sxi{2v^2}(56-8\sxi+2\xi-72\xi^{3/2}-13\xi^2+8\xi^{5/2}+3\xi^3)
  \ell_3\nl
  +\frac{2+\sxi}{2(1+\sxi)^2}(32+62\sxi+121\xi+122\xi^{3/2}+61\xi^2
  +10\xi^{5/2})I^{L_+}_0(0)\nl
  +\frac{16}{v^4}(3-2\sxi)I^{L_+}_{1-}(0)-16(1+\sxi)^4I^{L_+}_{1+}(0)\nl
  +\frac{(2-\sxi)\sxi}{2(1-\sxi)^2}(2+\sxi+10\xi-3\xi^{3/2}-6\xi^2)
  \left(I^{L_+}_{t-}(0)+I^{L_+}_{t+}(0)\right)\nl
  +\frac\sxi{2(1+\sxi)^2}(20+8\sxi+25\xi-4\xi^{3/2}-19\xi^2+2\xi^{5/2})
  I^{L_0}_0(0)\nl
  +\frac{(1-\sxi)\sxi}{(1+\sxi)^2}(4-20\sxi-3\xi+9\xi^{3/2}-2\xi^2)
  I^{L_0}_{1-}(0)\nl
  -4\sxi\left((2-\xi)\left(\hat I^{L_0}_{t-}(0)+\hat I^{L_0}_{t+}(0)\right)
  -4v(\ell_0^-+\ell_0^+)\right)\bigg],\\[12pt]
\lefteqn{\rho^{lt}_{12I}\ =\ N\sxi\bigg[
  -\frac\sxi{v}(20+19\sxi-28\xi-3\xi^{3/2})}\nl
  -\frac1{2v^2}(32-122\xi+28\xi^{3/2}+45\xi^2-4\xi^{5/2}-3\xi^3)\ell_3\nl
  +\frac1{2(1+\sxi)^2}(32+40\sxi+15\xi-26\xi^{3/2}-29\xi^2-8\xi^{5/2})
  I^{L_+}_0(0)
  +\frac{16}{v^4}(2-\sxi)I^{L_+}_{1-}(0)\nl
  +\frac\sxi{2(1-\sxi)^2}(8+17\sxi-26\xi-3\xi^{3/2}+8\xi^2)
  \left(I^{L_+}_{t-}(0)+I^{L_+}_{t+}(0)\right)\nl
  +\frac1{2(1+\sxi)^2}(16+40\sxi-5\xi-14\xi^{3/2}-\xi^2-4\xi^{5/2})
  I^{L_0}_0(0)\nl
  +\frac{(1-\sxi)\sxi}{(1+\sxi)^2}(8-5\sxi-3\xi+4\xi^{3/2})I^{L_0}_{1-}(0)\nl
  -4\left((2-\xi)\left(\hat I^{L_0}_{t-}(0)+I^{L_0}_{t+}(0)\right)
  -4v(\ell_0^-+\ell_0^+)\right)\bigg],\\[12pt]
\lefteqn{\rho^{lt}_{43A}\ =\ 2N\pi v\sxi(1+\xi),}\\[12pt]
\lefteqn{\rho^{lt}_{44A}\ =\ N\bigg[
  -\frac1{1+\sxi}(8+32\xi-5\xi^{3/2}-6\xi^2+3\xi^{5/2})}\nl
  +\frac2{v^2}(8-2\sxi+10\xi+\xi^{3/2}-\xi^2)\ell_1
  +\frac{4-\xi}{2(1-\sxi)}(8-16\sxi+8\xi+19\xi^{3/2}-3\xi^2)\ell_2\nl
  -\frac{v\sxi}{(1+\sxi)^2}(6-44\sxi-75\xi-42\xi^{3/2}-8\xi^2)\ell_3\nl
  -\frac\sxi{2(1+\sxi)^2}(16-16\sxi-70\xi-76\xi^{3/2}-35\xi^2-6\xi^{5/2})
  I^{L_+}_0(0)\nl
  +\frac{v\xi^{3/2}}{(1-\sxi)^2}(7-2\sxi-3\xi)
  \left(I^{L_+}_{t-}(0)-I^{L_+}_{t+}(0)\right)\nl
  +\frac\sxi{2(1+\sxi)^2}(16-8\sxi-56\xi-72\xi^{3/2}-37\xi^2-6\xi^{5/2})
  I^{L_0}_0(0)\nl
  -4v\sxi\left((2-\xi)\left(\hat I^{L_0}_{t-}(0)-I^{L_0}_{t+}(0)\right)
  -4v\ell_0^-\right)\bigg],\\[12pt]
\lefteqn{\rho^{ln}_{13I}\ =\ N\bigg[
  \frac{(1-\sxi)\sxi}{1+\sxi}(24+40\sxi-5\xi-3\xi^{3/2})}\nl
  -\frac2{1+\sxi}(8-10\sxi-16\xi-7\xi^{3/2})\ell_1\nl
  -\frac12(32-32\sxi+40\xi+20\xi^{3/2}-16\xi^2-3\xi^{5/2})\ell_2\nl
  +\frac{v\sxi}{(1+\sxi)^2}(2+53\xi+62\xi^{3/2}+16\xi^2)\ell_3\nl
  +\frac\sxi{2(1+\sxi)^2}(48+112\sxi+146\xi+108\xi^{3/2}+41\xi^2+6\xi^{5/2})
  I^{L_+}_0(0)\nl
  +\frac\sxi{2(1+\sxi)^2}(16+32\sxi-48\xi-88\xi^{3/2}-39\xi^2-6\xi^{5/2})
  I^{L_0}_0(0)\nl
  +\frac{(1+\sxi)\xi^{3/2}}{v}(5+3\sxi)
  \left(I^{L_+}_{t-}(0)-I^{L_+}_{t+}(0)\right)\nl
  -4v\sxi\left((2-\xi)\left(\hat I^{L_0}_{t-}(0)-I^{L_0}_{t+}(0)\right)
  -4v\ell_0^-\right)\bigg],\\[12pt]
\lefteqn{\rho^{ln}_{14I}\ =\ -2N\pi v\sxi(1+\xi),}\\[12pt]
\lefteqn{\rho^{ln}_{41A}\ =\ -2N\pi v^2\sxi,}\\[12pt]
\lefteqn{\rho^{ln}_{42A}\ =\ -2N\pi v^2\sxi,}\\[12pt]
\lefteqn{\rho^{ll}_{11U}\ =\ N\bigg[
  \frac2v(10-12\sxi-15\xi+20\xi^{3/2}-11\xi^2)}\nl
  +\frac1{v^2}(48+24\sxi-52\xi-60\xi^{3/2}+55\xi^2+12\xi^{5/2}-3\xi^3)
  \ell_3\nl
  +\frac1{(1+\sxi)^2}(32+54\sxi+78\xi+90\xi^{3/2}+71\xi^2+30\xi^{5/2}+5\xi^3)
  I^{L_+}_0(0)\nl
  +\frac{16}{v^4}(3-2\sxi+\xi)I^{L_+}_{1-}(0)-16(1+\sxi)^4I^{L_+}_{1+}(0)\nl
  +\frac\sxi{(1-\sxi)^2}(26-14\sxi-18\xi+\xi^{3/2}+6\xi^2+3\xi^{5/2})
  \tintpp\nl
  -\frac1{(1+\sxi)^2}(16+22\sxi+2\xi-34\xi^{3/2}-3\xi^2+14\xi^{5/2}-\xi^3)
  I^{L_0}_0(0)\nl
  +\frac{2(1-\sxi)\sxi}{(1+\sxi)^2}(10+8\sxi+10\xi+\xi^{3/2}-5\xi^2)
  I^{L_0}_{1-}(0)\nl
  +4(2-\xi)\tintsp\bigg],\\[12pt]
\lefteqn{\rho^{ll}_{11L}\ =\ N\bigg[
  -\frac1{2v}(96+32\sxi-50\xi-97\xi^2+3\xi^3)}\nl
  +\frac\sxi{4v^2}(32-40\sxi+88\xi-104\xi^{3/2}-24\xi^2+3\xi^{5/2}-3\xi^{7/2})
  \ell_3\nl
  -\frac1{2(1+\sxi)^2}(64+156\sxi+296\xi+340\xi^{3/2}+217\xi^2+70\xi^{5/2}
  +9\xi^3)I^{L_+}_0(0)\nl
  -\frac{16}{v^4}(3-2\sxi)I^{L_+}_{1-}(0)+16(1+\sxi)^4I^{L_+}_{1+}(0)\nl
  -\frac\sxi{2(1-\sxi)^2}(4+8\sxi+12\xi-25\xi^{3/2}-2\xi^2+7\xi^{5/2})
  \tintpp\nl
  -\frac\sxi{2(1+\sxi)^2}(4-24\sxi+4\xi-5\xi^{3/2}-6\xi^2+11\xi^{5/2})
  I^{L_0}_0(0)\nl
  -\frac{(1-\sxi)^2\sxi}{(1+\sxi)^2}(4-8\sxi+7\xi^{3/2})I^{L_0}_{1-}(0)\nl
  -\xi\tintsp\bigg],\\[12pt]
\lefteqn{\rho^{ll}_{12U}\ =\ N\bigg[
  \frac{4\sxi}v(16+\sxi-20\xi-\xi^{3/2})
  -\frac{2\sxi}{v^2}(4-16\xi+3\xi^{3/2}-3\xi^{5/2})\ell_3}\nl
  +\frac\sxi{(1+\sxi)^2}(48+134\sxi+166\xi+99\xi^{3/2}+24\xi^2+\xi^{5/2})
  I^{L_+}_0(0)\nl
  -\frac{16}{v^4}(1-2\sxi-\xi)I^{L_+}_{1-}(0)-16(1+\sxi)^4I^{L_+}_{1+}(0)\nl
  +\frac\xi{(1-\sxi)^2}(2+18\sxi-19\xi-4\xi^{3/2}+7\xi^2)\tintpp\nl
  -\frac\xi{(1+\sxi)^2}(6+22\sxi+\xi-12\xi^{3/2}-\xi^2)I^{L_0}_0(0)\nl
  +\frac{2(1-\sxi)^2\xi}{(1+\sxi)^2}(2-2\sxi-3\xi)I^{L_0}_{1-}(0)\nl
  +4\xi\tintsp\bigg],\\[12pt]
\lefteqn{\rho^{ll}_{12L}\ =\ N\bigg[
  \frac\xi{2v}(70+32\sxi-89\xi+3\xi^2)}\nl
  -\frac\sxi{4v^2}(32+104\sxi+104\xi-176\xi^{3/2}-40\xi^2+27\xi^{5/2}
  -3\xi^{7/2})\ell_3\nl
  +\frac\sxi{2(1+\sxi)^2}(32+156\sxi+232\xi+187\xi^{3/2}+82\xi^2+15\xi^{5/2})
  I^{L_+}_0(0)\nl
  -\frac{16}{v^4}(1+2\sxi-2\xi)I^{L_+}_{1-}(0)-16(1+\sxi)^4I^{L_+}_{1+}(0)\nl
  +\frac\xi{2(1-\sxi)^2}(4-24\sxi+5\xi+10\xi^{3/2}+\xi^2)\tintpp\nl
  +\frac\xi{2(1+\sxi)^2}(12+32\sxi-15\xi-18\xi^{3/2}+5\xi^2)I^{L_0}_0(0)\nl
  +\frac{(1-\sxi)\xi}{(1+\sxi)^2}(4+20\sxi+\xi-9\xi^{3/2})I^{L_0}_{1-}(0)\nl
  -\xi\tintsp\bigg],\\[12pt]
\lefteqn{\rho^{ll}_{43F}\ =\ -8N\pi v\xi,}\\[12pt]
\lefteqn{\rho^{ll}_{44F}\ =\ N\bigg[
  -\frac4{1+\sxi}(8+12\sxi+20\xi-9\xi^{3/2}-7\xi^2)}\nl
  +\frac4{v^2}(4+28\xi+8\xi^{3/2}-11\xi^2-5\xi^{5/2})\ell_1\nl
  +\frac2{1-\sxi}(24+8\sxi-6\xi+14\xi^{3/2}-\xi^2-3\xi^{5/2})\ell_2\nl
  +\frac{2v}{(1+\sxi)^2}(8+16\sxi+58\xi+66\xi^{3/2}+31\xi^2+6\xi^{5/2})
  \ell_3\nl
  -\frac\sxi{(1+\sxi)^2}(32+36\sxi-8\xi-46\xi^{3/2}-32\xi^2-7\xi^{5/2})
  I^{L_+}_0(0)\nl
  +\frac{2v\sxi}{(1-\sxi)^2}(8-6\sxi+2\xi^{3/2}-\xi^2)
  \left(I^{L_+}_{t-}(0)-I^{L_+}_{t+}(0)\right)\nl
  -\frac1{(1+\sxi)^2}(16+32\sxi+40\xi+28\xi^{3/2}+34\xi^2+28\xi^{5/2}+7\xi^3)
  I^{L_0}_0(0)\nl
  +8v\left((2-\xi)\left(\hat I^{L_0}_{t-}(0)-I^{L_0}_{t+}(0)\right)
  -4v\ell_0^-\right)\bigg].
\end{eqnarray}